\documentclass[10pt]{article}
\usepackage{graphicx}
\usepackage{amssymb}
\usepackage{amsmath}

\DeclareGraphicsExtensions{.jpg, .eps}
\title{Adjusted Bayesian inference for selected parameters}
\author{Daniel Yekutieli}

\newtheorem{thm}{Theorem}[section]
\newtheorem{cor}[thm]{Corollary}
\newtheorem{prop}[thm]{Proposition}
\newtheorem{lem}[thm]{Lemma}
\newenvironment{rem}{\refstepcounter{thm} \bigskip \par \noindent
    {\bf Remark \thethm}\ }{\bigskip \par}
\newenvironment{dfn}{\refstepcounter{thm} \bigskip \par \noindent
    {\bf Definition \thethm}\ }{\bigskip \par}
\newenvironment{exa}{\refstepcounter{thm} \bigskip \par \noindent
        {\bf Example \thethm}\ }{\bigskip \par}
\newenvironment{pf}{\medskip \par \noindent {\it Proof.}\ }{\hfill \P \bigskip \par}

\begin{document}
\baselineskip=15pt
\textwidth=27pc

\maketitle

\begin{abstract}
We address the problem of providing inference from a Bayesian perspective
for parameters selected after viewing the data.
We present a Bayesian framework for providing inference for selected parameters,
based on the observation that providing Bayesian inference for selected parameters is a truncated data problem.
We show that if the prior for the parameter is non-informative,
or if the parameter is a ``fixed'' unknown constant,
then it is necessary to adjust the Bayesian inference for selection.
Our second contribution is the introduction of Bayesian False Discovery Rate controlling methodology,
which generalizes existing Bayesian FDR methods that are only defined in the two-group mixture model.
We illustrate our results by applying them to simulated data and data from
a microarray experiment.
\end{abstract}


\section{Introduction}
We discuss providing Bayesian inference for parameters selected after viewing the data.
Current thought is that selection has no affect on the inference of parameters from a Bayesian perspective.
We show that this is not necessarily the case. Consider generating a sample from a Bayesian framework
by randomly generating the parameter and conditional on the parameter data are generated.
In one case, selection is applied to samples of the parameter and the data,
and in the other case the parameter is sampled and then selection is applied to data samples.
The example below shows that selection matters in the latter case, but not in the former case.

\begin{exa} \label{exa-00}
Let $\theta$ denote students' true academic ability.
The marginal density of $\theta$ in the population of high school students is $N(0, 1)$.
The observed academic ability of students in high school is $Y \sim N( \theta, 1)$,
and students with $0 < Y$ are admitted to college.
We wish to predict a student's true academic ability from his observed academic ability --
but only if the student is admitted to college.
We will show that the Bayesian inference is different for a random high school student
than for a random college student.

We first consider the case of a college professor predicting $\theta$ for a student in his class.
The joint distribution of $(\theta, Y)$ for a random college student can be generated
by generating $(\theta, y)$ for a random high school student and selecting
$(\theta, y)$ only if $0 < y$.
Thus the joint density of $(\theta, y)$ used for predicting $\theta$ is
\begin{equation} \label{eq89a}
f_S ( \theta, y)  \;  \propto  \;
e^{ - \frac{  \theta^2 }{2}} \cdot e^{ - \frac{ (\theta - y)^2}{2}}  /  \Pr ( Y > 0)
\;  \propto \; e^{  - \frac{ (\theta - y/2)^2}{ 2 \cdot (1/2)}},
\end{equation}
and the conditional distribution of $\theta$ given $Y = y$ is  $N( y/2 , 1/2)$.
The predicted academic ability for a student with $y = 1$ is  $E ( \theta | \; y = 1)= 0.5$.

For the case of the high school teacher predicting $\theta$ for a student in his class,
we assume that there is a high school regulation instructing teachers
to predict academic ability only for students that can be admitted to college.
This means that for any true academic ability $\theta$,
the values of $Y$ used to predict $\theta$
are drawn from the $N(\theta, 1)$ density truncated by the event $0 < Y$.
Since $\theta$ for a random student is $N(0,1)$,
the joint density of $(\theta, y)$ used for predicting $\theta$ is
\begin{equation} \label{eq89b}
f_S ( \theta, y)  \; \propto \;
 e^{ - \frac{  \theta^2 }{2}} \cdot e^{ - \frac{ (\theta - y)^2}{2}}  / \Pr( Y > 0 |  \theta).
\end{equation}
In this case there is no closed expression for the conditional distribution of $\theta$ given $Y = y$,
but since $\Pr( Y > 0 |  \theta)$ decreases in $\theta$ then it is stochastically smaller than $N( y/2 , 1/2)$,
and the predicted academic ability for a student with $y = 1$ is $E ( \theta | \; y = 1) = 0.10$.
\end{exa}

In this paper, we address selection that arises in the statistical analysis of large data sets
in which the aim is to  find interesting parameters and then  provide
inferences for these selected parameters.
Throughout the paper we use the following simulated example to illustrate the discussion.
One can consider it as an example of a microarray experiment in which $\theta_i$
is the log-fold change in expression of Gene $i$
and $Y_i$ is the observed log expression ratio.
We will now show that even when the selection is applied to the parameter and the data,
it is necessary to correct Bayesian inference for selection if the prior on the parameter
is non-informative.

\begin{exa} \label{exa1}
The simulation includes $10^5$ iid samples of $(\theta_i, Y_i)$.
To generate $\theta_i$,  we first sample $\lambda_i$ from $\{ 10, 1 \}$ with probabilities $0.90$ and $0.10$,
and then draw $\theta_i$ from the laplace distribution,
$\pi_1 (\theta_i | \lambda_i ) = \lambda_i \cdot \exp(- \lambda_i \cdot | \theta_i |) / 2$.
Thus the marginal distribution of $\theta_i$ is
\begin{equation} \label{eqn90}
\pi (\theta_i) \; = \;  0.9 \cdot \pi_1( \theta_i | \lambda_i = 10) + 0.1 \cdot \pi_1( \theta_i | \lambda_i = 1).
\end{equation}
$Y_i = \theta_i + \epsilon_i$, with $\epsilon_i$ independent $N(0,1)$.


In our analysis we apply the level $q=0.2$ Benjamini and Hochberg (1995) (hereafter BH) FDR controlling
procedure to the two sided p-values, $p_i = 2 * \{ 1 - \Phi (|Y_i|)\}$, in order to find interesting $\theta_i$,
and then construct $0.95$ credible interval for each interesting $\theta_i$.
The BH procedure yielded $R = 932$ discoveries ($p_{(932)} = 0.001862 < 0.001864 =  0.2 \cdot 932 / 10^5$):
the set of $\theta_i$ with $|Y_i| > 3.111$. The $932$ selected $(\theta_i, Y_i)$ are displayed in Figure 1.

We use two prior models for constructing credible intervals for $\theta_i$.
In the first model the prior distribution for $\theta_i$ is $\pi(\theta_i)$ in (\ref{eqn90}).
In this case the posterior distribution of $\theta_i$ (we derive it later in the paper)
is the conditional distribution of $\theta_i$ given $Y_i$.
Thus the probability that $\theta_i$ is in the $0.95$ credible interval constructed for it is, per definition, $0.95$.
Furthermore, since for each $(\theta_i, Y_i)$ we draw a new value of $\theta_i$ from $\pi(\theta_i)$,
selection should have no affect on the Bayesian inference.
And indeed, $0.953$ of the selected $\theta_i$ ($888$ out of $932$) are covered by their respective $0.95$ credible intervals.

In the second model we assume that the marginal distribution of $\theta_i$ is unknown
and we replace it with the non-informative prior, $\pi(\theta_i) = 1$.
The posterior distribution of $\theta_i$ for this prior distribution is $N(Y_i, 1)$.
Thus $Y_i \pm Z_{1 - 0.05/2}$ is a $0.95$ credible interval for $\theta_i$ (these are the red lines in Figure 1).
Even though the posterior distribution for non-informative priors is not the conditional distribution
of the parameters given the data, these are equal tail credible intervals based on minimally-informative priors
known to provide good frequentist performance (Carlin and Louis, 1996, Section $4.3$) that are expected
to cover approximately $0.95$ of the $\theta_i$.
These credible intervals cover $0.951$ of all $100,000$ $\theta_i$,
but only $0.654$ of the selected $\theta_i$ ($610$ out of $932$).
\end{exa}

Before presenting our inferential framework in Section 1.6,
we review a frequentist approach for discovering interesting parameters and providing inferences for these discoveries in Section 1.1.
In section 1.2 we further motivate the importance of our problem by  reviewing  literature on providing inference for interesting parameters in genomic studies.
In Section 1.3-1.5 several aspects of Bayesian analysis that are relevant to our work are reviewed.

\subsection{Control over the false coverage-statement rate}
Soric (1989) asserted that the goal of many scientific experiments is to discover non-zero effects,
and as a result made the important observation that it is mainly the discoveries that are reported and included into science,
and warned that unless the proportion of false discoveries in the set of declared discoveries is kept small
there is danger that a large part of science is untrue.

BH considered the problem of testing $m$ null hypotheses $H_1 \cdots H_m$,
of which $m_0$ are true null hypotheses.
They referred to the rejection of a null hypothesis as a discovery
and the rejection of a true null hypothesis as a false discovery.
To limit the occurrence of false discoveries when testing multiple null hypotheses
BH introduced the False Discovery Rate $FDR = E \{ V  / \max( R, 1) \}$,
where $R$ is the number of  discoveries and $V$ is the number of false discoveries,
and introduced the BH multiple testing procedure that controls the FDR at a nominal level $q$.

Benjamini and Yekutieli (2005) generalized the Benjamini and Hochberg testing framework.
In their selective inference framework there are $m$ parameters
$\theta_1 \cdots \theta_m$, with corresponding estimators $T_1 \cdots T_m$,
and the goal is to construct valid marginal confidence intervals (CIs) for the
subset of parameters
selected by a given selection rule ${\cal S} (t_1 \cdots t_m) \subseteq \{ 1 \cdots m \}$.
They showed that CIs constructed for selected parameters no longer ensure the nominal coverage
probability, and suggested the False Coverage-statement Rate (FCR) as
the appropriate criterion to capture the error for CIs constructed for selected parameters.
The FCR is also defined by $E \{ V  / \max( R, 1) \}$, however $R$ is the number of CIs constructed
and $V$ is the number of non-covering CIs.
Benjamini and Yekutieli (2005) introduced a method of ensuring $FCR \le q$ for independent $T_1 \cdots T_m$
and any selection criterion:
construct marginal $1 - R \cdot q / m$ CIs for each of the $R$ selected parameters.
In cases where each $\theta_i$ can be associated with a null value $\theta^0_i$
and the selection criteria are multiple testing procedures that test $\theta_i = \theta^0_i$ vs. $\theta_i \ne \theta^0_i$,
Benjamini and Yekutieli (2005) showed that the level $q$ BH procedure can be expressed
as the least conservative multiple testing procedure that ensures that all level
$q$ FCR adjusted CI for $\theta_i$, for which the null hypothesis is rejected,
will not cover the respective $\theta^0_i$.
Furthermore, they show that for independent $T_1 \cdots T_m$ if all $\theta_i \ne \theta^0_i$ then applying the level $q$ BH procedure
to select the parameters and declaring each selected $\theta_i$ greater than $\theta^0_i$ if $T_i > \theta^0_i$
and smaller than $\theta^0_i$ if $T_i < \theta^0_i$ controls the directional FDR
(expected proportion of selected parameters assigned the wrong sign)
at level $q / 2$.

\begin{exa} \label{exa1a}
Note that in example 1.2 all $\theta_i \ne 0$,
thus for any multiple testing procedure $FDR \equiv 0$.
However declaring  $\theta_i$ positive for the BH discoveries with $Y_i > 0$ and negative for the BH discoveries with $Y_i < 0$,
ensures directional-FDR less than $0.1$.
The number of simulated positive selected $\theta_i$ with negative $Y_i$
and negative selected $\theta_i$ with positive $Y_i$ is $56$,
thus the observed directional-FDR is $0.060$.

The red lines in Figure 1 are two-sided Normal $0.95$ CIs:
$Y_i \pm Z_{1 - 0.05 / 2}$ (recall that these are also the non-informative prior $0.95$ credible intervals
from Example 1.2).
These $0.95$ CIs cover $95,089$ of the $100,000$ simulated $\theta_i$,
but only $610$ of the $932$ selected $\theta_i$,
thus the observed FCR is $0.346$.
The green lines are $0.05$ FCR-adjusted CIs:
$Y_i \pm Z_{1 - 0.05 \cdot 932 / ( 2 \cdot 10^5)}$.
The observed FCR for the FCR-adjusted CIs is $0.046$.
\end{exa}

\subsection{Selective inference in Genomic association studies}
The need to correct inference for selection is widely recognized in Genome-wide association studies (GWAS).
GWAS typically test association between a disease and hundreds of thousands of markers located throughout the human genome,
often expressed as an odds ratio of manifesting the disease in carriers of a risk allele.
Only multiplicity-adjusted significant findings are reported.
This limits the occurrence of false positives, however it introduces bias into the odds ratio estimates.
Analyzing $301$ published studies covering 25 different reported associations,
Lohmueller et al. (2003) found that for 24 associations the odds ratio in the first
positive report exceeded the genetic effect estimated by meta-analysis
of the remaining studies.
Zollner and Pritchard (2007) suggest correcting for the selection bias by providing point estimates and CIs based on
the likelihood conditional on having observed a significant association.
Zhong and Prentice (2008) further assume that in the absence of selection the log odds ratio estimator is
Normally distributed. Similarly to our Bayesian analysis of the simulated example,
they base their inference on a truncated normal conditional likelihood.

\subsection{Parameter selection in Bayesian analysis}
Berry and Hochberg (1999) comment that the Bayesian treatment of the multiplicity problem also includes
decision analysis, rather than just finding posterior distributions.

Scott and Berger (2006) discuss Bayesian analysis of microarray data.
The prior model for $\theta_i$, the expectation of the log-fold change in expression of Gene $i$,
is that $\theta_i = 0$ with probability $p$ and $\theta_i \sim N(0, V)$ with probability $1-p$.
The decision analysis performed in Scott and Berger (2006) is the discovery of the subset of active genes.
Scott and Berger (2006) declare a gene active ($\theta_i \ne 0$) if the posterior
expected loss of this action is smaller than the posterior expected loss of declaring the gene
inactive ($\theta_i = 0$).
The loss function for deciding that $\theta_i = 0$ is proportional to $|\theta_i|$,
and the loss for erroneously deciding that $\theta_i \ne 0$ is the fixed cost of
doing a targeted experiment to verify that the gene is in fact active.

The decision analysis in Bayesian FDR analysis of microarray data is also deciding which genes are active.
In Efron et al. (2001), $\theta_i$ is selected if its local FDR, which is the posterior probability given $y_i$ that $\theta_i = 0$,
is less than a nominal value $q$.
Storey (2002, 2003) suggests specifying selection rules for which the positive FDR (pFDR), defined
as the conditional probability that $\theta_i = 0$ given that $\theta_i$ is selected, is less than $q$.
In the optimal discovery procedure suggested in Storey (2007), the statistic used for specifying the selection rule
is a plug-in estimator of the local FDR. Storey (2007) shows that the optimal discovery procedure provides
the maximal probability of selecting $\theta_i$ among all selection rules with the same pFDR level.

\subsection{Selection bias in Bayesian analysis} \label{sect1}
Dawid (1994) explains why selection should have no effect on Bayesian inference:
``Since Bayesian posterior distributions are already fully conditioned on the data,
the posterior distribution of any quantity is the same, whether
it was chosen in advance or selected in the light of the data.''
%
%

Senn (2008) reviews the disparity between Bayesian and frequentist approaches regarding selection.
He considers the example of providing inference for  $\theta_{i^*}$,
the effect of the pharmaceutical associated with the largest sample mean $y_{i^*}$,
among a class of $m$ compounds with $Y_i \sim N(\theta_i, 4)$.
He first shows that if $\theta_i$ are iid $N(0,1)$
the posterior distribution of  $\theta_{i^*}$ is $N(y_{i^*} / 5, 4/5)$.
He then assumes a hierarchical model in which the treatments form a compound class.
The class effect is $\lambda \sim N(0, 1 - \gamma^2)$ and $\theta_i$ are iid $N(\lambda, \gamma^2)$.
In this case he shows that the posterior distribution of $\theta_{i^*}$
depends on the number of other compounds and their overall mean, however it is
unaffected by the fact that $\theta_{i^*}$ was selected because it corresponds to the largest sample mean.

The observation that Bayesian inference may be affected by selection
was already made in Mandel and Rinott (2007, 2009).
Mandel and Rinott (2007) consider the scenario of providing inference for $p$, the probability of success
in a binomial experiment, conditional on observing two or more successes. Similar to Example \ref{exa-00},
they distinguish between the case that in each binomial experiment $p$ is drawn independently from its prior
distribution and the case that the value of $p$ is the same in all binomial experiments, and they show that in
the second case the Bayesian inference is affected by selection.

\subsection{Fixed and random effects in Bayesian analysis}
In the Bayesian framework there can be no fixed effects since the parameters are regarded as having probability distributions.
However, discussing one-way classification Box and Tiao (1973, Section 7.2) use the sampling theory terminology of
fixed and random effects to distinguish between situations in which the individual means can be regarded as distinct values expected to bear no
strong relationship to each other that can take on values anywhere within a wide range,
and situations in which the individual means can be regarded as draws from a distribution.
Box and Tiao illustrate this distinction with the example of one-way classification of several groups of laboratory yields.
In the first case the groups correspond to different methods of making a particular chemical product,
while in the second case the groups correspond to different batches made by the same method.
The distinction only carries through to the prior model elicited for the group means.
In the first case the group means are elicited flat non-informative priors.
They call this model the fixed effect model.
In the second case the group means are iid $N( \lambda, \sigma^2)$.
This model is called the random effect model.

\subsection{Preliminary definitions} \label{prel-dfn}
Let $\theta$ denote the parameter, $Y$ denote the data and $\Omega$ is the sample space of $Y$.
$\pi (\theta)$ is the prior distribution of $\theta$, and $f(y | \theta)$ is the likelihood function.
The multiple parameters for which inference may or may not be provided, are actually multiple functions
of $\theta$: $h_1 (\theta), h_2 (\theta), \ldots$ . In selective inference for each $h_i (\theta)$ there is
a subset $S^i_{\Omega} \subseteq \Omega$, such that inference is provided for
$h_i (\theta)$ only if $y \in S^i_{\Omega}$ is observed.
For example,  in our analysis of microarray data in Section 6,
$Y$ is the entire set of observed gene expression levels;
$\theta = ( \sigma^2, \mu)$ consists of the variances and expectations of the
log-expression levels for all the genes in the array; and inference is provided for
$h_g (\theta) = \mu_g$, the expectation of the log-fold change in expression of Gene $g = 1 \cdots G$,
only if Gene $g$ is declared differentially expressed by the BH procedure.

Control over the FCR is a frequentist mechanism for providing selective inference.
Notice that in Example \ref{exa1} a randomly selected $\theta_i$ is covered by its FCR-adjusted CI with probability $ \ge 0.95$.
But this frequentist selective inference mechanism suffers from several intrinsic limitations:
it is impossible to incorporate prior information on the parameters;
it does not provide selection adjusted point estimates or selection-adjusted inference for functions of the parameters;
the selection adjustment is the same regardless of the selection criterion applied
and the value of the estimator.
Figure $1$ suggests that the selection adjustment needed shrinks the CIs toward $0$,
rather than just widening the CIs; and the the larger $| Y_i|$ the smaller selection adjustment is needed for $\theta_i$.


In selective inference the entire data set $Y=y$ is observed.
However, as inference is provided for $h_i (\theta)$ only if $y \in S^i_{\Omega}$,
then $Y=y$ used for providing selective inference for $h_i (\theta)$ is actually a realization of the
joint distribution of $(\theta, Y)$, truncated by the event that $y \in S^i_{\Omega}$
(describing Bayesian selective inference a truncation problem was suggested by Bradley Efron in private communication;
for a discussion on truncation see Mandel (2007) and Gelman et al. (2004) Section 7.8).
Thus in order to provide Bayesian selective inference for $h_i (\theta)$
we introduce a framework for providing Bayesian inference based on the truncated distribution of $(\theta, Y)$.
We call this inference selection-adjusted Bayesian (saBayes) inference

Predicting true academic ability from observed academic ability
for a high school student and for a college student, discussed in Example \ref{exa-00},
are Bayesian selective inference problems in which inference is provided for
$h(\theta) = \theta$ only if $S_{\Omega} = \{ y : 0 < y \}$ occurs.
Notice that even though the selection mechanism is different,
in both cases, $(\theta, y)$ for which $\theta$ is predicted from $y$ are truncated samples from the
distribution of $(\theta, Y)$ in the population of all high school students.

\subsection{Outline of the paper}
In Section 2 we discuss modelling selection-adjusted Bayesian inference:
we provide an operative definition for the joint truncated distribution of $(\theta, Y)$;
we distinguish between parameters according to the way their distribution is affected by selection,
and derive the joint truncated distribution of $(\theta, Y)$ in either case;
for either case, and also for parameters with non-informative priors,
we define the components (i.e., prior, likelihood and posterior) of saBayes inference;
we then specifically derive these components for $(\theta, Y)$ that correspond to
Box and Tiao's random effect model and fixed effect model.
In Section 3 we define saBayes inference as the Bayes rules in Bayesian selective inference.
We also present a Bayesian FCR for the random effect model and explain the relation
between saBayes inference and providing FCR control.

In Section 4 we present Bayesian FDR controlling methodology for specifying selection rules in
the random effects model for cases in which selection is used for making statistical discoveries.
We also provide an eBayes algorithm for applying this methodology
in cases that correspond to Box and Tiao's fixed effect model.
In Section 5 we explain the relation between the Bayesian FDR methods presented in Section 4
and existing Bayesian FDR methods,
and describe how to provide saBayes inference in the two group mixture model.

In Section 6 we analyze microarray data. The goal of the analysis
is to find over-expressed and under-expressed genes while controlling directional FDR $\le 0.05$,
and to provide inference for the change in expression for these selected genes.
The level $0.10$ BH procedure applied to $t$ statistic p-values
fails to discover any differentially expressed genes.
Applying the level $0.10$ BH procedure to p-values corresponding to
hybrid frequentist/eBayes moderated t-statistics yields 245 discoveries,
however it is not clear how to provide frequentist selective inference for these discoveries.
For comparison, our level $0.05$ Bayesian FDR selection rule based on the moderated t-statistic
yields $1124$ discoveries, and the level $0.05$ Bayesian FDR selection rule
based on the optimal statistic yields $1271$ discoveries.
In the second part of the analysis,
we provide Bayesian selective inference for the expected log2-fold change in expression
for a differentially expressed gene.

The paper concludes with a discussion of the conceptual and methodological contributions of this paper.

\section{Modelling selection-adjusted Bayesian inference}
The primary problem in modelling saBayes inference is specifying the joint truncated distribution of $(\theta, Y)$,
which we denote $f_S (\theta, y)$.
It is important to note that $f_S (\theta, y)$ is the joint distribution of $(\theta, Y)$
according to which selective inference is provided for $h(\theta)$,
and not the joint distribution of $(\theta, Y)$,
$f(\theta, Y) = \pi(\theta) \cdot f( y | \theta)$.
We use this characterization for defining $f_S (\theta, y)$.

\begin{dfn} \label{dfn345}
Assume that selective inference for $h(\theta)$ involves an action $\delta(Y)$ associated with a loss function
$L( h(\theta), \delta)$. $f_S (\theta, y)$ is defined as the distribution over which the expected loss
\begin{equation} \label{avg-loss}
r_S ( \delta) =  \int_{\theta} \int_{y \in S_{\Omega}}   f_S ( \theta, y ) \cdot   L (h(\theta), \delta(y))  \ dy d\theta
\end{equation}
is the average risk incurred in selective inference for $h (\theta)$.
\end{dfn}

\subsection{``Fixed,'' ``random,'' and ``mixed'' parameters in Bayesian selective inference}
Example \ref{exa-00} illustrated that $f_S (\theta, y)$ is determined by the way selection acts on $\theta$.
Unlike Box and Tiao who use the terms fixed and random effects to describe the type of prior distribution elicited for $\theta$,
we use the terms ``fixed,'' ``random,''  and ``mixed'' parameters to describe the way the distribution of
$\theta$ is affected by selection.
For each parameter type, we derive $f_S (\theta, y)$,
$\pi_S (\theta)$ the marginal truncated distribution of $\theta$,
and $f_S (y | \theta)$ the truncated conditional distribution of $Y | \theta$.


\subsubsection{The ``fixed'' parameter truncated sampling model}
We call $\theta$ a ``fixed'' parameter if its distribution is unaffected by selection and selection
is applied to the conditional distribution of $Y$ given $\theta$.
''Fixed'' parameters are unknown constants whose values are assumed to be sampled from $\pi(\theta)$ and remain unchanged.
Thus for each value of $\theta$,
the risk incurred in providing selective inference for $h(\theta)$
is the expected loss over the truncated conditional distribution of $Y | \theta$
\[
 \int_{y \in S_{\Omega}}  f(y | \theta)  / Pr(S_{\Omega} | \; \theta) \cdot L (h(\theta), \delta(y))  \  d y,
\]
for $ Pr(S_{\Omega} | \; \theta) =   \int_{y \in S_{\Omega}}  f(y | \theta) \ d y$,
and the average risk is its expectation over the marginal density of $\theta$
\begin{equation}
r_S (\delta) =  \int_{\theta} \int_{y \in S_{\Omega}}  \pi(\theta) \cdot f(y | \theta) / Pr(S_{\Omega} | \; \theta)
\cdot L (h(\theta), \delta(y)) \ d y  \ d \theta. \label{fixd-risk}
\end{equation}
Thus in this case the joint truncated distribution of $(\theta, Y)$ is
\begin{equation} \label{fixd-joint}
f_S ( \theta, y ) \;  = \; I_{S_\Omega} (y) \cdot \pi(\theta) \cdot f ( y | \; \theta) / \Pr(S_\Omega | \;
\theta),
\end{equation}
the marginal truncated density of $\theta$ is
\begin{equation} \label{dens29ab}
\pi_S (\theta) = \pi(\theta),
\end{equation}
and the truncated conditional distribution of  $Y | \theta$ is
\begin{equation} \label{cond-lik}
f_S ( y | \theta) \;  = \; I_{S_\Omega} (y) \cdot f ( y | \; \theta) / \Pr(S_\Omega | \; \theta).
\end{equation}

\subsubsection{The ``random'' parameter truncated sampling model}
We call $\theta$ a ``random'' parameter in cases where selection is applied to the joint distribution of $(\theta, Y)$.
In this case $\theta$ is drawn from $\pi(\theta)$
and $Y$ is drawn from $f(y | \theta)$, but inference is provided for $h(\theta)$ only
for $(\theta, y)$ with $y \in S_{\Omega}$.
Thus the average risk incurred in providing selective inference $h(\theta)$ is
\begin{equation} \label{rand-risk}
r_S (\delta) =  \int_{\theta} \int_{y \in S_{\Omega}}  \pi(\theta) \cdot f(y | \theta) / Pr(S_{\Omega})
\cdot L (h(\theta), \delta(y)) \ d y  \ d \theta,
\end{equation}
for $ Pr(S_{\Omega} ) =   \int_{\theta} \int_{y \in S_{\Omega}} \pi(\theta) \cdot  f(y | \theta) \ d y$.
Thus the truncated distribution of $(\theta, Y)$ is
\begin{equation} \label{rand-joint}
f_S ( \theta, y ) \;  = \; I_{S_\Omega} (y) \cdot \pi(\theta) \cdot f ( y | \; \theta) / \Pr(S_\Omega ).
\end{equation}
Integrating out $y$ yields the marginal truncated distribution of $\theta$
\begin{equation} \label{dens5b}
\pi_S (\theta)  =  \pi(\theta) \cdot \Pr(S_\Omega | \;  \theta) / \Pr(S_{\Omega} ).
\end{equation}
Dividing (\ref{rand-joint}) by (\ref{dens5b}) reveals that in this case the
truncated distribution of $Y |\theta$ is also the conditional likelihood in (\ref{cond-lik}).

\subsubsection{The ``mixed'' parameter truncated sampling model}
We call $\theta$ a ``mixed'' parameter in cases where selection is applied to the conditional distribution
of $(\theta, Y)$ given $\lambda$, for a hyperparameter $\lambda \sim \pi_2(\lambda)$ with
$\theta | \lambda \sim  \pi_1 (\theta | \lambda)$.
Thus conditioning on $\lambda$, $\theta$ is ``random'' and the average risk
incurred in providing selective inference is
\begin{equation} \label{mixd-cond-risk}
\int_{\theta} \int_{y \in S_{\Omega}}  \pi_1(\theta | \lambda) \cdot f(y | \theta) / Pr(S_{\Omega} | \lambda )
\cdot L (h(\theta), \delta(y)) \ d y  \ d \theta,
\end{equation}
where $Pr(S_{\Omega} | \lambda ) = \int_{\theta} \int_{y \in S_{\Omega}} \pi_1(\theta | \lambda) \cdot f(y | \theta) \ d y \ d \theta$.
Taking expectation over $\lambda$ yields the average risk
\begin{equation} \label{mixd-risk}
r_S (\delta) =   \int_{\lambda} \int_{\theta} \int_{y \in S_{\Omega}}  \pi_2 (\lambda) \cdot \frac{\pi_1(\theta | \lambda)
\cdot f(y | \theta)}{Pr(S_{\Omega} | \lambda )}
\cdot L (h(\theta), \delta(y)) \ d y  \ d \theta \ d \lambda.
\end{equation}
Thus the truncated density of $(\lambda, \theta, y)$ is
\begin{equation} \label{dens-lambda}
f_S (\lambda, \theta, y) = I_{S_\Omega} (y) \cdot \pi_2 (\lambda) \cdot \pi_1(\theta | \;  \lambda) \cdot f(y | \theta)
 / \Pr(S_{\Omega} | \;  \lambda).
\end{equation}
Changing the order of integration in (\ref{mixd-risk}) we get
\begin{equation} \label{mixd-risk2}
r_S (\delta) =   \int_{\theta} \int_{y \in S_{\Omega}}
\{  \int_{\lambda} \frac{ \pi_2 (\lambda) \cdot \pi_1(\theta | \lambda) }{Pr(S_{\Omega} | \lambda )} d \lambda \}
\cdot f(y | \theta) \cdot L (h(\theta), \delta(y)) \ d y  \ d \theta,
\end{equation}
and thus the truncated density of $(\theta, y)$ is
\begin{equation} \label{dens33b}
f_S (\theta, y) = I_{S_\Omega} (y) \cdot f(y | \;  \theta) \cdot  \int_{\lambda}
 \pi_2(\lambda) \cdot \pi_1(\theta | \;  \lambda) / \Pr(S_{\Omega} | \;  \lambda) \ d\lambda.
\end{equation}
Integrating out $y$ yields the marginal truncated distribution of $\theta$
\begin{equation} \label{dens33c}
\pi_S (\theta)   \; = \;  \Pr(S_\Omega | \;  \theta)   \int_{\lambda} \frac{ \pi_2(\lambda)  \pi_1(\theta | \;  \lambda) }
{\Pr(S_{\Omega} | \;  \lambda)}  d\lambda.
\end{equation}
And again, dividing (\ref{dens33b}) by (\ref{dens33c}) reveals that
the truncated distribution of $Y | \theta$ is $f_S( y |  \theta)$
in (\ref{cond-lik}).

\begin{rem} \label{rem11}
It is important to note that classifying $\theta$ a ``fixed'', ``random'', or ``mixed'' parameter
is context dependent and must be done on a case by case basis.
In Example \ref{exa-00},  $\theta$ is an unknown constant,
for both a random college student and a random high school student.
However, comparing Expressions (\ref{eq89a}) and (\ref{eq89b}) with  (\ref{fixd-joint}) and (\ref{rand-joint}),
reveals that $\theta$ is a ``fixed'' parameter for a random high school student,
and a ``random'' parameter for a random college student.

Senn's example of providing inference for the most active compound
can be expressed as a selective inference problem in which, for $i = 1 \cdots m$,
inference is provided for $h_i (\theta) = \theta_i$ only if $S^i_{\Omega} = \{ y : y_i = max (y_1 \cdots y_m)  \}$ occurs.
When $\theta$ is the vector of treatment effects of $m$ distinct compounds, each component of $\theta$ is a distinct unknown constant
whose value is sampled from $N(\lambda, \gamma^2)$ and remains unchanged, therefore $\theta$ is a ``fixed'' parameter.
Now suppose that $\theta_i \sim N(\lambda, \gamma_i)$ are batch effects of $m$ batches treated by
a single compound, with compound effect $\lambda \sim N(0, 1 - \gamma^2)$.
In this case, $\lambda$ is a ``fixed'' unknown constant, and conditional on $\lambda$,
$\theta$ is a ``random'' batch effect. Thus $\theta$ is a ``mixed'' parameter.
\end{rem}

\subsection{Defining the components of Bayesian selective inference}
The selection-adjusted prior distribution is, when it is available,
the marginal truncated distribution of $\theta$.
We have shown that the selection adjusted prior distribution for ``fixed'', ``random'' or ``mixed'' $\theta$
is $\pi_S (\theta)$ given in (\ref{dens29ab}), (\ref{dens5b}) or (\ref{dens33c}).
Note that to specify the marginal truncated distribution of $\theta$,
we need $\pi (\theta)$ to be the marginal distribution of $\theta$
and we need to know how selection acts on $\theta$.

An important case in which $\pi(\theta)$ is not the marginal distribution of $\theta$
is when $\pi(\theta)$ is a non-informative prior distribution.
Non-informative priors are used to allow conditional analysis on $\theta$
when no prior information on $\theta$ is available (Berger 1985, Section 3.3.1).
As $Y$ also provides all the information on $\theta$ in the truncated data problem,
we argue that the prior distribution used for saBayes inference should also be a non-informative prior.
We further argue that while the lack of prior knowledge on $\theta$ may affect our decision to provide selective inference,
the opposite is not true -- the decision to provide inference only for certain values of $Y$ should have no effect
on the non-informative prior elicited for $\theta$.  We therefore suggest using the same non-informative
prior for saBayes inference, $\pi_S (\theta) =  \pi (\theta)$.
Which means that if the prior for $\theta$ is non-informative then it is treated as a ``fixed'' parameter.

The selection adjusted likelihood is $f_S ( y |  \theta)$ in (\ref{cond-lik}),
the truncated conditional distribution of $Y$ given  $\theta$.
Note that conditioning on $\theta$ ensures that the selection adjusted likelihood
is the same in the three truncated sampling models and does not depend on the marginal distribution of $\theta$.

The selection-adjusted posterior distribution is defined by
\begin{equation} \label{saBayes-post}
\pi_S (\theta | \; y) \; = \;  \pi_S (\theta) \cdot  f_S ( y | \; \theta) / m_S (y),
\end{equation}
for $m_S (y) \; = \;  \int  \pi_S (\theta) \cdot  f_S ( y | \; \theta) d \theta$.
For non-informative priors it is generated by updating the non-informative prior according to
the selection-adjusted likelihood.
For ``fixed'', ``random'' or ``mixed'' $\theta$ it is the truncated conditional distribution
of $\theta | Y$. Thus $\pi_S (\theta | y) \propto f_S (\theta, y )$.
But note that only for ``random'' $\theta$, for which $ f_S (\theta, y) \propto f (\theta, y)$,
the selection-adjusted posterior distribution is unaffected by selection.

\begin{rem} \label{rem-Dawid}
Dawid argues that selection has no effect on posterior distributions
since conditioning on the selection event is made redundant by conditioning on $Y = y$.
Note that this only applies for the case of ``random'' $\theta$, for which selection can be expressed
as conditioning on an event $S$ in the sample space of $(\theta, Y)$.
Hence, as Dawid argues, for $(\theta, y) \in S$ the truncated posterior distribution
is the same as the untruncated posterior distribution:
\[
\pi_S (\theta | y) = \pi ( \theta | S, Y = y) = \frac{ f( \theta, S, Y =y)}{ f(S, Y =y)} = \frac{ f( \theta, Y  = y)}{ f( Y =y)}
=  \pi ( \theta | Y = y) = \pi (\theta | y).
\]
Whereas for ``fixed'' and ``mixed'' $\theta$, for which selection cannot generally be expressed as conditioning on
an event in the sample space of $(\theta, Y)$, $\pi_S (\theta | \; y)$
is generally different than $\pi (\theta | \; y)$ as demonstrated in Example \ref{exa-00} and in Example \ref{exa-0}.
We illustrate how this point applies to our simulated data in Example \ref{exa3333}.
\end{rem}

\begin{exa} \label{exa-0}
Senn (2008) concludes that selection has no effect on the Bayesian inference
because in his analysis $\theta$ is a ``random'' parameter.
In Remark \ref{rem11} we suggest that in this kind of analysis
$\theta$ will most likely be a  ``fixed'' or a ``mixed'' parameter.
We therefore compute the selection-adjusted posterior mean of $h_2 (\theta) = \theta_2$ for $m=2$ and $y = (0, 2)$,
for ``mixed'' and ``fixed'' $\theta$.

However, as $S_{\Omega}^2 = \{ ( \theta, y) : y_2 \ge y_1 \}$, then  $\Pr( S^2_{\Omega}  | \lambda) \equiv \Pr( S^2_{\Omega} )  = 0.5$,
and the ``mixed'' parameter model truncated joint density defined in (\ref{dens33b})
reduces to the ``random'' parameter joint density in (\ref{rand-joint}).
Thus in this case, also for ``mixed'' $\theta$,  the conditional distribution of $\theta_2$ is unaffected by selection.
We use Expression (4) in Senn (2008) to compute the conditional mean of $\theta_2$.
For  $\gamma^2 = 1$ it equals $0.4$ and for $\gamma^2 = 0.5$ it equals $0.384$.

The selection-adjusted joint density of $\theta$ for ``fixed'' $\theta$ is given by
\[
\pi_S (\theta_1, \theta_2  | \; y = (0, 2) ) \; \propto \; \frac{
e^{ - \frac{ \lambda^2}{2 \gamma^2}}
\cdot e^ {  - \frac{ (\theta_1 - \lambda)^2}{ 2 \cdot (1 - \gamma^2)}}
\cdot e^ {  - \frac{ (\theta_2 - \lambda)^2}{ 2 \cdot (1 - \gamma^2)}}
\cdot e^ {  - \frac{ (0 - \theta_1)^2}{ 2 \cdot 4}}
\cdot e^ {  - \frac{ (2 - \theta_2)^2}{ 2 \cdot 4}}}{ Pr ( Y_2 \ge Y_1 |  \theta_1, \theta_2) }.
\]
In this case the selection adjustment increases the posterior distribution of $\theta$ values with $\theta_2 < \theta_1$, thereby stochastically
decreasing the marginal posterior distribution of $\theta_2$.
For $\gamma^2 = 1$ the conditional mean of $\theta_2$ is $0.164$ and for $\gamma^2 = 0.5$ it is $0.257$.
\end{exa}

\subsection{Modeling Bayesian selective inference in the random effect model}
Using the terminology suggested by Box and Tiao, we call the model for $\theta = (\theta_1 \cdots \theta_m) $
and $Y = \{ Y_1 \cdots Y_m \}$, where $\theta_i$ are iid $\pi (\theta_i)$
and $Y_i | \theta_i$ are independent $f(y_i | \theta_i)$,
a random effect model.

In the random effect model $\theta$ can be a ``random'' parameter, a ``fixed'' parameter,
and even a ``mixed'' parameter when there are iid ``fixed'' $\lambda_i$ for which $\theta_i | \lambda_i$
are independent ``random'' parameters. In any case the joint distribution of $(\theta, Y)$ is
\begin{equation} \label{re0}
f(\theta, y) = \pi(\theta) \cdot f(y | \theta) = \Pi_{i = 1}^{m}  \pi (\theta_i) \cdot \Pi_{i = 1}^{m} f(y_i | \theta_i).
\end{equation}
In selective inference for $h_i(\theta) = \theta_i$
with $S^i_{\Omega} = \{ y : y_i \in S_{marg} \}$, incorporating (\ref{re0}) into (\ref{fixd-joint})
yields the  ``fixed'' $\theta$ selection adjusted joint distribution of $(\theta, Y)$
\begin{eqnarray}
\lefteqn{f_S (\theta, y)  =   I_{S^i_\Omega} (y) \cdot  \Pi_{j = 1}^{m}  \{ \pi (\theta_j) \cdot f(y_j | \theta_j)\} /
 \Pr( S^i_{\Omega} | \theta)} \nonumber \\
 &  = &   \Pi_{j \ne i} \{  \pi (\theta_j) f( y_j | \theta_j) \}
  I_{S_{marg}}(y_i)\  \pi(\theta_i)  f(y_i | \theta_i)  / \Pr( Y_i \in S_{marg}  | \theta_i). \label{eqn45}
\end{eqnarray}
Integrating out  $\theta^{(i)}$ and $y^{(i)}$ in (\ref{eqn45})
yields the selection adjusted distribution of $(\theta_i, Y_i)$ for ``fixed'' $\theta$
\begin{equation} \label{re-67}
f_S (\theta_i, y_i) =   I_{S_{marg}}(y_i) \cdot \pi(\theta_i) \cdot f(y_i | \theta_i) / \Pr( Y_i \in S_{marg}  | \theta_i).
\end{equation}
Similarly, incorporating (\ref{re0}) into (\ref{rand-joint}) and integrating out $\theta^{(i)}$ and $y^{(i)}$,
yields the selection adjusted joint distribution of $(\theta_i, Y_i)$ for ``random'' $\theta$
\begin{equation} \label{re2}
f_S (\theta_i, y_i) =  I_{S_{marg}}(y_i) \cdot \pi(\theta_i) \cdot f(y_i | \theta_i) / \Pr(Y_i \in S_{marg}).
\end{equation}
Incorporating (\ref{re0}) into (\ref{dens33b}) and integrating out  $\theta^{(i)}$ and $y^{(i)}$,
yields the ``mixed'' $\theta$  selection adjusted distribution of $(\theta_i, Y_i)$
\begin{equation} \label{re3}
f_S (\theta_i, y_i) =  I_{S_{marg}}(y_i) \cdot f(y_i | \;  \theta_i) \cdot \int
\frac{ \pi_2(\lambda_i) \cdot \pi_1(\theta_i | \;  \lambda_i)}
{\Pr(Y_i \in S_{marg} | \lambda_i)} d\lambda_i.
\end{equation}

\subsubsection{The non-exchangeable random effect model}  \label{par-sel}
The non-exchangeable random effect model is a generalization of the random effect
model for situations in which $\theta_i$ are distinct values expected to bear no
strong relationship one to each other, i.e. situations for which Box and Tiao would suggest the fixed effect model.
In the non-exchangeable random effect model
$\theta_i$ are independent but have distinct prior distributions, $\pi^i (\theta_i)$, while
$Y_i | \theta_i$ are still independent $f (y_i | \theta_i)$.
Thus the joint distribution of $(\theta, Y)$ is
\begin{equation} \label{non-exc-rf}
f (\theta, y) = \pi(\theta) \cdot f(y | \theta) = \Pi_{i = 1}^{m}  \pi^i (\theta_i) \cdot \Pi_{i = 1}^{m} f (y_i | \theta_i).
\end{equation}
The marginal distribution of $(\theta_i, Y_i)$ is
\[ f (\theta_i, y_i) = \pi^i (\theta_i) \cdot f (y_i | \theta_i). \]
But in selective inference for $h_i(\theta) = \theta_i$ with $S^i_{\Omega} = \{ y : y_i \in S_{marg} \}$,
the selection adjusted joint distribution of $(\theta_i, Y_i)$ for ``fixed'' $\theta$ is
\begin{equation} \label{re1-exc}
f_S (\theta_i, y_i) =  I_{S_{marg}}(y_i) \cdot \pi^i (\theta_i) \cdot f (y_i | \theta_i) / \Pr( Y_i \in S_{marg}  | \theta_i).
\end{equation}

\begin{exa} \label{exa3333}
Notice that $(\theta, Y)$ in Example \ref{exa1} are generated by the random effect model
that the components of $\theta = (\theta_1 \cdots \theta_{100,000})$ are independently
drawn from $\pi(\theta_i)$ in (\ref{eqn90})
and $Y_i | \theta_i$ are independent $f( y_i | \theta_i) = \phi(y_i - \theta_i)$.
Figure $1$ is a scatter plot of $932$ $(\theta_i, y_i)$ with $|y_i| > 3.111$;
Figure $4$ displays the $470$ components with $y_i > 3.111$.
For comparison, in the comparable non-exchangeable random effect model:
for $i = 1 \cdots 90,000$, $\theta_i \sim \pi_1( \theta_i | \lambda_i = 10)$
and for $i = 90,001 \cdots 100,000$, $\theta_i \sim \pi_1( \theta_i | \lambda_i = 1)$.

It is important to note that in Example 1.2 we draw a single realization from the joint untruncated distribution of $(\theta, Y)$.
To observe the difference between ``random'', ``fixed'' and ``mixed'' $\theta$ we conduct another set of simulations,
in which we sample $1000$ realizations of $(\theta, Y)$ from its truncated distributions for
$h_1(\theta) = \theta_1$ with $S^1_{\Omega} = \{ y : |y_1| > 3.111 \}$ for ``random'', ``fixed'' and ``mixed'' $\theta$.
Each realization from the ``random'' $\theta$ truncated distribution is generated by repeatedly sampling $(\theta, Y)$
from its untruncated distribution, keeping the first $(\theta, y)$ for which $|y_1| > 3.111$.
To generate each realization from the ``fixed'' $\theta$ truncated distribution, we sample $\theta$ from $\pi(\theta)$ and then
repeatedly sample $Y$, keeping the first $y$ with $|y_1| > 3.111$.
 As the components of $(\theta, Y)$ are independent the distribution of $(\theta_2,  \cdots, \theta_{100,000},  Y_2,  \cdots, Y_{100,000})$
is the same in the three truncation models.
Figure $2$ displays the scatter plots of the $Y_1 > 3.111$ realizations of $(\theta_1, Y_1)$ for each truncation model.
The left panel is the scatter plot for the ``random'' $\theta$ model.
In this case the joint density of $(\theta_1, Y_1)$, given in (\ref{re2}), is
 \[
 \pi ( \theta_1) \cdot \phi( y_1  - \theta_1 ),
 \]
and it is identical to the joint density of $(\theta_i, Y_i)$ displayed in Figures $4$ and
the distribution of $(\theta_i, Y_i)$ for $Y_i > 3.111$ in Figure $1$.
The right panel is the scatter plot for the ``fixed'' $\theta$ model.
In this case the joint density of $(\theta_1, Y_1)$, given in (\ref{re-67}), is
\[
 \pi ( \theta_1) \cdot \phi( y_1  - \theta_1 ) /  \Pr( |Y_1| > 3.111 \ | \theta_1).
 \]
Comparing the right and left panels reveals that in this model, for each value of $Y_1$, the conditional distribution
$\theta_1$ is shrunk towards $0$.
To generate each realization from the ``mixed'' $\theta$ truncated distribution,
for $i = 1 \cdots 100,000$ we independently sample $\lambda_i$
from $\{ 10, 1 \}$, with probabilities $0.90$ and $0.10$, and then we repeatedly sample $(\theta, Y)$,
$\theta_i \sim \pi_1 ( \theta_i | \; \lambda_i )$ and $Y_i \sim \phi(\theta_i)$, keeping the first $(\theta, y)$ for which $|y_1| > 3.111$.
The joint density of $(\theta_1, Y_1)$ given in (\ref{re3}) is
\[
\{ \frac{ 0.9 \cdot \pi_1 ( \theta_1 | \lambda_1 = 10 )}{ \Pr( |Y_1| > 3.111 \ | \lambda_1=10)} +
 \frac{   0.1 \cdot \pi_1 ( \theta_1 | \lambda_1 = 1 )} { \Pr( |Y_1| > 3.111 \ | \lambda_1=1)} \}
 \cdot \phi( y_1  - \theta_1 ).
 \]
Comparing the three panels of Figure 2
reveals that in this model the shrinking of the distribution of $\theta_1 | Y_1 = y_1$
towards $0$ is weaker than in the ``fixed'' $\theta$ model.
\end{exa}

\section{Selection-adjusted Bayesian inference}
To define saBayes inference,
we express the average risk incurred by providing selective inference for $h(\theta)$
\begin{eqnarray}
r_S ( \delta)
& = & \int_{\theta} \int_{y \in S_{\Omega}}  L (h(\theta), \delta(y)) \cdot  \pi_S (\theta) \cdot f_S (y | \theta) dy d\theta  \nonumber \\
& = &   \int_{y \in S_{\Omega}} [ \int_{\theta}
 L (h(\theta), \delta(y)) \cdot \pi_S (\theta | \; y) d \theta ] \cdot  m_S(y) dy.\label{saBayes-risk}
\end{eqnarray}
Thus the Bayes rules in selective inference are the actions minimizing the
selection-adjusted posterior expected loss
\[
\rho_S ( \delta, y ) \; = \; \int L (h(\theta), \delta(y)) \cdot \pi_S (\theta | y) d\theta,
\]
and in general Bayesian selective inference should be based on
the selection-adjusted posterior distribution of $h(\theta)$,  $\pi_S (h(\theta) | \; y)$.
Selection-adjusted $1-\alpha$ credible intervals for $h(\theta)$
are subsets $A$ for which $\Pr_{\pi_S (h(\theta) | \; y)} ( h(\theta) \in A) = 1-\alpha$,
and the posterior mean or mode of $\pi_S (h(\theta) | \; y)$ can serve as selection-adjusted
point estimators for $h(\theta)$.

\begin{exa} \label{exa4400}
We provide saBayes inference for the data simulated in Example \ref{exa1}
for two selected parameters:
$h_{12647} (\theta) = \theta_{12647}$ with $S^{12647}_{\Omega} = \{ y: |y_{12647}| > 3.111 \}$,
and $h_{90543} (\theta)$ $= \theta_{90543}$ with $S^{90543}_{\Omega}$ $= \{ y: |y_{90543}| > 3.111 \}$.
Since we have drawn $\theta$ from $\pi(\theta)$ and $Y$ from $f( y | \theta)$ then $\theta$ is a ``random'' parameter.
Recall that we use two prior models for $\theta$ in our analysis.
In the first model we assume that  $(\theta, Y)$ was generated by a random effect model
with $\pi(\theta_i)$ in (\ref{eqn90}).
In this model the saBayes posterior distribution of $\theta_i$
is proportional to the distribution of $(\theta_i, Y_i)$ in (\ref{re2})
\begin{equation} \label{eq333}
\pi_S ( \theta_i | \; y_i )  \; \propto \; \pi (\theta_i) \cdot \phi( y_i  - \theta_i ).
\end{equation}
In the second model $(\theta, Y)$ is generated by a non-exchangeable random effect
model with unknown $\pi^i(\theta_i)$
(note that if it were assumed that $\theta$ was generated by a random effect model then
eBayes could be used to estimate $\pi(\theta_i)$).
Thus, following Box and Tiao, we use the flat non-informative prior $\pi^i(\theta_i) = 1$ in our analysis.
The flat prior unadjusted posterior distribution of $\theta_i$ is
\begin{equation} \label{eq333a}
\pi ( \theta_i  \  | \;  y_i)  \; \propto \; \phi( y_i  - \theta_i ).
\end{equation}
The non-informative prior saBayes posterior distribution of $\theta_i$ is proportional to the
distribution of $(\theta_i, Y_i)$ for ``fixed'' $\theta$ in (\ref{re-67})
\begin{equation} \label{pppp}
\pi_S( \theta_i | \;  y_i)  \; \propto \; \phi( y_i  - \theta_i )  / Pr( S_{marg} | \; \theta_i ),
\end{equation}
with $Pr( S_{marg} | \; \theta_i) = \Phi( -3.111 - \theta_i) +  1-  \Phi( 3.111 - \theta_i)$.

Figure $3$ displays the posterior distributions of $\theta_{12647}$ (left panel) and $\theta_{90543}$ (right panel).
The flat prior unadjusted posterior mean and mode of $\theta_{12647}$ equal $Y_{12647} = 3.40$, and the $0.95$ credible interval is $[1.44, 5.36]$.
The saBayes posterior distribution of $\theta_{12647}$ is shrunk towards $0$.
The ``random'' $\theta$ saBayes posterior distribution of $\theta_{12647}$ is bimodal with a spike at $0$ and a mode at $2.40$,
the posterior mean is $1.68$, and the $0.95$ credible interval is $[-0.11, 4.20]$.
The flat prior saBayes posterior mode of $\theta_{12647}$ is $0.74$, the posterior mean is $1.88$,
and the $0.95$ credible interval is $[-0.04, 4.64]$.

The flat prior unadjusted posterior mean and mode of $\theta_{90543}$ equal $Y_{90543} = 5.59$,
and the $0.95$ credible interval is $[3.63, 7.55]$.
The much larger $Y_{90543}$ produces a non-negligible likelihood only for $\theta_i$ values that correspond
to almost certain selection. Thus in this case the selection adjustment is small:
the flat prior saBayes posterior mode is $5.57$, the posterior mean is $5.48$,
and the $0.95$ credible interval is $[3.26, 7.52]$.
The shrinking towards $0$ in the ``random'' $\theta$ model posterior is stronger:
the posterior mean and mode are $4.59$ and the $0.95$ credible interval is $[2.62, 6.55]$.
\end{exa}

\begin{rem}
It is important to note that as extremely unlikely values of $\theta$ with an extremely small selection probability can
have a large selection-adjusted likelihood,
the selection adjustment posterior distribution can be be very different than the unadjusted posterior distribution.
The selection-adjusted likelihood can even be non-informative and improper --
if the selection rule only includes the observed value $Y = y$ then the selection-adjusted likelihood is constant for all parameter values.
Example \ref{ex89} illustrates this phenomenon,
shows how it is affected by the choice of the selection rule and that it is not unique to Bayesian selective inference.
In this paper we employ selection rules whose selection probability is minimized at $\theta=0$
and approaches $1$ for large $|\theta|$, thus the selection adjustments shrink the likelihood towards $0$.
\end{rem}

\begin{exa} \label{ex89}
We derive the non-informative prior saBayes posterior distribution of $\theta_{12647}$,
given in (\ref{pppp}), for an alternative one-sided selection rule
$S^{12647}_{\Omega} = \{ y : y_{12647} > 3.111 \}$.
In this case the selection-adjusted posterior is stochastically smaller and much more diffuse.
The selection-adjusted posterior mode is $0.19$ and the selection-adjusted posterior mean is $-2.87$;
the $0.95$ selection-adjusted credible interval is $[-15.41, 3.91]$.
An unlikely value $\theta_{12647} = -5.87$, with unadjusted likelihood
$\phi(-5.87 - 3.40) = 8.73 \times 10^{-20}$ and selection probability $\Phi(-5.87 - 3.111) = 1.34 \times 10^{-19}$,
has the same selection-adjusted posterior density as the unadjusted posterior mode $\theta_{12647} = 3.40$,
i.e. $\pi_S ( \theta_{12647} = -5.87 | \ Y_{12647} = 3.40 ) = \pi_S ( \theta_{12647} = 3.40 | \ Y_{12647} = 3.40 )$.

We now show that frequentist selection adjusted inference can also be very different than the unadjusted frequentist inference,
and highly dependent on the type of selection rule used.
The flat prior unadjusted $0.95$ credible interval for  $\theta_{12647}$, $[1.44, 5.36]$ is also a $0.95$ frequentist
confidence interval for  $\theta_{12647}$.
To construct selection-adjusted frequentist $0.95$ confidence intervals for $\theta_{12647}$
we begin by testing, at level $0.05$ and for each value of $\theta_0$,
the null hypothesis that $\theta_{12647} = \theta_0$.
The sampling distribution of $Y_{12647} | \theta_{12647} = \theta_0$ is $f_S ( y_{12647} | \theta_{12647})$
in (\ref{cond-lik}) with $\theta_{12647} = \theta_0$.
Thus we reject the null hypothesis that $\theta_{12647} = \theta_0$
if $y_{12647}$ is smaller than the $0.025$ quantile or larger than the $0.975$ quantile of $f_S ( y_{12647} | \theta_0)$,
and the $0.95$ confidence interval for $\theta_{12647}$ is the set of $\theta_0$ values
for which the null hypothesis  that $\theta_{12647} = \theta_0$ is not rejected for $y_{12647} = 3.40$.
For the selection rule $S^{12647}_{\Omega} = \{ y : | y_{12647} | > 3.111 \}$
the $0.95$ confidence interval for $\theta_{12647}$ is $[-0.37, 5.03]$.
While for $S^{12647}_{\Omega} = \{ y :  y_{12647}  > 3.111 \}$ the $0.95$ confidence interval for $\theta_{12647}$
is $[-9.44, 5.03]$.
\end{exa}

\subsection{FCR control in the random effect model} \label{sec-fcr}
We define the FCR for $(\theta, Y)$ generated by the random effect model.
The initial set of parameters is $\theta_1 \cdots \theta_m$.
The subset of selected parameters is $\{ \theta_i : y_i \in S_{marg} \}$,
and a marginal confidence interval $A_{marg} (y_i)$ is constructed for each selected $\theta_i$.
For $i = 1 \cdots m$, let $R_i = I(Y_i \in S_{marg})$ and $V_i = I (Y_i \in S_{marg}, \theta_i \notin A_{marg} (Y_i) )$.
$R = \sum R_i$ is the number of selected parameters,
$V = \sum V_i$ is the number of non-covering confidence intervals,
and $FCP = V / max(1, R)$ is the false coverage-statement proportion.
In Benjamini and Yekutieli (2005) FCR refers to a frequentist FCR
that corresponds to $E_{Y | \theta} FCP$ for $(\theta, Y)$ generated by a random effect model.
In this paper FCR is a Bayesian FCR, defined by $E_{\theta , Y } FCP$.
We also consider the positive FCR, $pFCR = E_{\theta, Y } ( FCP | R>0)$.

\subsubsection{Relation between FCR control and Bayesian selective inference}
Note that for $i = 1 \cdots m$, the indicators $R_i$ and $V_i$ are defined for
the joint (untruncated) distribution of $(\theta, Y)$.
The event $R_i = 1$ is given by $\{ (\theta, y) : y_i \in S_{marg} \}$.
The conditional distribution of $(\theta, Y)$ given $R_i = 1$ is
\begin{equation} \label{re3ab}
f (\theta, y | R_i = 1)  =  I_{S_{marg}} (y_i) \cdot \Pi_{j = 1}^m
\{ \pi (\theta_j) \cdot f(y_j | \theta_j) \} / \Pr(Y_i \in S_{marg}),
\end{equation}
and integrating out $\theta^{(i)}$ and $y^{(i)}$ yields the conditional distribution
of $(\theta_i, Y_i)$ given $R_i = 1$ to be
\begin{equation} \label{re2ab}
f (\theta_i, y_i | R_i = 1)  =  I_{S_{marg}} (y_i) \cdot \pi (\theta_i) \cdot f(y_i | \theta_i) / \Pr(Y_i \in S_{marg}).
\end{equation}
This is the same as the ``random'' parameter selection-adjusted distribution of $(\theta_i, Y_i)$ given in (\ref{re2}).
This implies that the conditional probability that the confidence interval constructed for $\theta_i$ fails to cover
$\theta_i$, given that $\theta_i$ is selected, can be expressed as the average risk incurred in
selective inference for $h_i(\theta) = \theta_i$ with
$S^i_{\Omega} = \{ y : y_i \in S_{marg} \}$ and with $\theta$ being a ``random'' parameter,
for the loss function
$L(\theta_i, A_i(y)) = I ( \theta_i \notin A_{marg}(y_i))$:
\begin{eqnarray}
\lefteqn{\Pr( V_i = 1 | R_i = 1) \;  = \; } \nonumber \\
&& \int_{\theta_i}  \int_{y_i \in S_{marg}}
\frac{ \pi(\theta_i) f(y_i | \theta_i) \cdot I (\theta_i \notin A_{marg} (y_i) )}{\Pr(Y_i \in S_{marg})} d y_i  d \theta_i
\; = \; r_S. \label{re2abc}
\end{eqnarray}
$\Pr( V_i = 1 | R_i = 1, Y_i = y_i)$
is equal to the ``random'' $\theta$ selection adjusted posterior expected loss
\begin{equation} \label{re2ad}
\rho (y_i) \; = \; \int  I (\theta_i \notin A_{marg} (y_i))
\cdot \pi_S (\theta_i | y_i ) d \theta_i,
\end{equation}
for $\pi_S (\theta_i | y_i )  \propto \pi(\theta_i) \cdot f( y_i | \theta_i )$
the ``random'' $\theta$ selection adjusted posterior distribution.

\begin{prop} \label{prop1}
The pFCR and $EV / ER$ are equal to the ``random'' $\theta$ average risk in (\ref{re2abc}).
If $A_{marg}(y_i)$ are $1 - \alpha$ credible intervals for $\theta_i$ based
on the ``random'' $\theta$ selection adjusted posterior distribution then $pFCR = \alpha$.
\end{prop}

\begin{pf}
In the random effect model $\{ V_i : R_i = 1 \}$ are mutually independent
with $\Pr( V_i = 1 | R_i = 1) = r_S$.
Thus for each value of $R=k$, $V \sim Binom(k, r_S)$,
and conditioning on $R > 0$ yields $pFCR =  r_S $.
Note that the numerator and denominator in (\ref{re2abc}) equal $E V_i$ and $E R_i$.
Thus $EV / ER  = EV_i / ER_i$ is also $r_S$.
Lastly, for $1-\alpha$ selection-adjusted credible intervals based
on $\pi_S (\theta_i | y_i)$,
$r_S = \rho ( y_i) \equiv \alpha$.
\end{pf}

\begin{rem} \label{rem-FCR}
We have shown that in the random effect model,
regardless of whether $\theta$ is ``random'', ``fixed'' or ``mixed'',
the pFCR equals the ``random'' $\theta$ selection-adjusted average risk.
As $\hbox{pFCR} \ge \hbox{Bayesian-FCR}$
the ``random'' $\theta$ average risk can serve as a conservative estimate for Bayesian-FCR.
In particular, for large $R$ the sampling dispersion of FCP and of $V / E R$ is small,
thus the FCP, Bayesian-FCR, frequentist-FCR, and pFCR that equals $E V / E R$,
which we discuss in the context of specifying selection rules in the non-exchangeable random effect model, are almost the same.
\end{rem}

\begin{rem} \label{rem-FCR-noninf}
Recall that if $\pi(\theta_i)$ is a noninformative prior then the selection adjusted posterior
distribution for ``random'' $\theta$ is defined
\begin{equation} \label{exp-87}
\pi_S ( \theta_i | y_i) \propto
\pi(\theta_i) \cdot f(y_i | \theta_i) / \Pr( S_{marg} | \theta_i ).
\end{equation}
As credible intervals based on non-informative priors are expected to provide approximate coverage probability,
when $\pi (\theta_i)$ is a non-informative prior then $1 - \alpha$ credible intervals based on
$\pi_S( \theta_i | y_i)$ in (\ref{exp-87}) yield $\rho (y_i) \approx \alpha$.
Thus Proposition \ref{prop1} implies that for non informative priors the ``fixed'' $\theta$
marginal $1-\alpha$ credible intervals yield approximate level $\alpha$ FCR control.
\end{rem}

\begin{exa} \label{exa444a}
Figure $4$ displays $(\theta_i, y_i)$ generated in Example \ref{exa1} with $y_i > 3.111$.
The red and green dashed curves are the $0.95$ confidence intervals from Figure $1$.
The red curves also correspond to the $0.95$ credible intervals for $\theta_i$ for the flat prior unadjusted posterior (\ref{eq333a}).
The blue curves are the $0.95$ saBayes credible intervals for the flat prior selection-adjusted posterior
in (\ref{pppp}), and the light blue curves are the $0.95$ saBayes credible intervals for the ``random''
$\theta$ selection-adjusted posterior in (\ref{eq333}).

According to Proposition \ref{prop1} the pFCR for ``random'' $\theta$ $0.95$ saBayes credible intervals
constructed for selected $(\theta_i, y_i)$ is $0.05$.
In Example 1.2 we have seen that the FCP for these credible intervals for the $932$ selected $\theta_i$ was $0.047$.
As the flat prior unadjusted credible intervals are $0.95$ frequentist confidence intervals,
we expect the coverage proportion for all $100,000$ $\theta_i$ to be close to $0.95$.
We have seen that these CIs cover $95,089$ of the $100,000$ $\theta_i$,
but that the FCP for the $932$ selected parameters is $0.346$.
Benjamini and Yekutieli (2005) explain this phenomenon from a frequentist perspective.
Remark \ref{rem-FCR-noninf} offers a Bayesian explanation: in order to provide approximate
FCR control for non informative priors the credible intervals should be based on
the ``fixed'' $\theta$ selection adjusted posterior in (\ref{pppp}),
rather than the ``random'' $\theta$ selection adjusted posterior in (\ref{eq333a}).
And indeed, the FCP of the credible intervals based on (\ref{pppp}) is $0.040$.
\end{exa}

\section{Specifying FDR controlling selection rules in the random effect model}
We will now present Bayesian methodology for specifying selection rules in the random effect model and the
non-exchangeable random effect model
for cases in which selection is applied for making statistical discoveries.
Similarly to the BH FDR controlling approach,
we seek to control the proportion of false discoveries committed.
Unlike BH, in which discoveries refer to rejection of null hypotheses
and the statistics used for specifying the selection rule are p-values testing these null hypotheses,
in our approach any event in the parameter space can be considered a discovery
and any statistic may be used for specifying the selection rule.
But, as suggested in Storey (2007), we will show that for any given discovery
the optimal statistic is the posterior probability that the discovery is false.

As in Section \ref{sec-fcr}, we assume that $(\theta, Y)$ are generated by the random effect model;
$\theta_i$ is selected if $y_i \in S_{marg}$;
and the inference provided for $\theta_i$ if it is selected is declaring that $\theta_i \in A_{marg}(y_i)$.
However now $A_{marg}(y_i)$ is an event that corresponds to making a statistical discovery regarding $\theta_i$.
For example, in the microarray analysis in Section 6, in which the discovery is declaring a gene either
over or under expressed, for $y_i > 0$ the discovery event is  $A_{marg} (y_i) = \{ \theta_i : \theta_i > 0 \}$.

Once declaring $\theta_i \in A_{marg}(y_i)$ corresponds to making a statistical discovery,
$R$ becomes the number of discoveries,
$V$ becomes the number of false discoveries,
$V / max(1, R) = FDP$ is the false discovery proportion, and $FCR = FDR$.
Thus Proposition \ref{prop1} yields the following result.
\begin{cor} \label{cor0-a}
In the random effect model the $pFDR$ equals $r_S$ in ($\ref{re2abc})$,
which is the conditional probability given that $\theta_i$ is selected
that the discovery regarding $\theta_i$ is false,
and  $\rho(y_i)$ in (\ref{re2ad}) is the conditional probability given selection and
given $Y_i = y_i$ that the discovery is false.
\end{cor}

Thus in order to ensure level $q$ FDR control,
when considering selection rules of the form $S_{marg} = \{ y_i : T(y_i) \le s \}$,
we suggest choosing $s$ for which  $r_S$ in ($\ref{re2abc})$ is $ \le q$.
Furthermore, reexpressing $r_S$
 \begin{eqnarray}
r_S & = &
\frac{ \int_{y_i \in S_{marg}} m(y_i) \cdot \int_{\theta_i}  \pi_S (\theta_i | y_i)
 \cdot I (\theta_i \notin A_{marg} (y_i) )  \ d \theta_i \ d y_i}
{{\Pr(Y_i \in S_{marg})}} \nonumber \\
& = &
\frac{ \int_{y_i \in S_{marg}} m(y_i) \cdot \rho(y_i) \ d y_i}
{ \int_{y_i \in S_{marg}} m(y_i) \ d y_i}
\end{eqnarray}
where $m (y_i) =  \int  \pi (\theta_i) \cdot f( y_i | \theta_i) d \theta_i$,
yields the following Neyman-Pearson Lemma type result, presented in Storey (2007).
\begin{cor} \label{cor1}
The selection rule of the form
$S_{marg} = \{ y_i : \rho ( y_i) \le s \}$ has the largest selection probability
of all selection rules with the same pFDR.
\end{cor}
Another option is to use $\rho (y_i)$ to directly specify the selection rule, by defining
\begin{equation} \label{loc-FDR}
 S_{marg} = \{ y_i : \rho ( y_i) \le q \}.
\end{equation}

\paragraph{}
Notice that unlike the continuum of possible credible intervals that can be constructed for $\theta_i$,
the number of possible discoveries that can be made regarding $\theta_i$ is
usually finite, e.g. discovering that $\theta_i$ is either negative or positive
or discovering that $\theta_i$ is the largest component in $\theta$.
In particular, when there is only a single possible discovery for all selected values of $y_i$, i.e. $A_{marg} (y_i) \equiv A_{marg}$,
then expressing the ``random'' $\theta$ average risk corresponding to this discovery
\begin{eqnarray}
r_S & =  & \int \int_{y_i \in S_{marg}}  I (\theta_i \notin A_{marg}) \cdot
\frac{  \pi (\theta_i) \cdot f (y_i | \theta_i)}{\Pr(Y_i \in S_{marg})} d y_i d \theta_i  \nonumber \\
& =  & \int  I (\theta_i \notin A_{marg}) \cdot
\frac{  \pi (\theta_i) \Pr(Y_i \in S_{marg} | \theta_i ) }{\Pr(Y_i \in S_{marg})} d \theta_i  \nonumber \\
& =  & \int  I ( \theta_i \notin A_{marg}) \cdot \pi_S (\theta_i) d \theta_ i, \label{pFDR=saprior}
\end{eqnarray}
for $\pi_S (\theta_i) =  \pi (\theta_i) \cdot \Pr(S_{marg} | \;  \theta_i) / \Pr(S_{marg})$
the ``random'' $\theta$ selection-adjusted prior density derived in (\ref{dens5b}), yields the following result.

\begin{cor} \label{cor0}
If $A_{marg} (y_i) \equiv A_{marg}$ then the $pFDR$ is equal to the
``random'' $\theta$ selection-adjusted prior probability that $\theta_i \notin A_{marg}$.
\end{cor}

\subsection{Specifying FDR controlling selection rules in the non-exchangeable random effect model}
In this subsection, $(\theta, Y)$ is generated by the non-exchangeable random effect model,
$\theta_i$ is selected if $y_i \in S_{marg}$,
and the inference provided for selected $\theta_i$ is the discovery that $\theta_i \in A_{marg}(y_i)$.
Let $A^1_{marg} \cdots A^D_{marg}$ denote the $D$ possible discoveries that can be made on $\theta_i$.
For $d = 1 \cdots D$, let $R^d$ denote the number of discoveries of $A^d_{marg}$
and let $V^d$ denote the number of false discoveries of $A^d_{marg}$.
The results in this section are derived under the assumption that $A_{marg} (y_i) \equiv A_{marg}$.
However as $ER = ER^1 + \cdots + ER^D$ and $EV = EV^1 + \cdots + EV^D$,
they can be easily extended for the case of $D>1$.

To derive the results in this section, we assume that there also exists
$(\tilde{\theta}, \tilde{Y})$, generated by the random parameter model that
$\tilde{\theta}_i$ are iid $\tilde{\pi} (\theta_i) = \sum_{i=1}^m \pi^i (\theta_i) / m$,
and $\tilde{Y}_i | \tilde{\theta}_i$ are independent $f(\tilde{y}_i | \tilde{\theta}_i)$.

\begin{lem} \label{lemm0}
For any subset $B$, $W_i = I (y_i \in S_{marg}, \theta_i \notin B)$,
and $\tilde{W}_i = I (\tilde{y}_i \in S_{marg}, \tilde{\theta}_i \notin B)$
\[ E \sum_{i=1}^m  W_i = E \sum_{i=1}^m  \tilde{W}_i. \]
\end{lem}

\begin{pf}
\begin{eqnarray}
E  \sum_{i=1}^m  W_i & = &  \sum_{i=1}^m  \Pr ( Y_i \in S_{marg}, \theta_i \notin B)  \nonumber \\
&  = &  \sum_{i=1}^m \int_{\theta_i \notin B}  \int_{y_i \in S_{marg}} \pi^i (\theta_i) \cdot f( y_i | \theta_i)  dy_i d \theta_i   \nonumber \\
&  = &  \sum_{i=1}^m \int_{\theta_1 \notin B}  \int_{y_1 \in S_{marg}} \pi^i (\theta_1) \cdot f( y_1 | \theta_1)  dy_1 d \theta_1   \nonumber \\
&  = &  m \cdot  \int_{\theta_1 \notin B}  \int_{y_1 \in S_{marg}} \sum_{i=1}^m \pi^i (\theta_1) / m \cdot f( y_1 | \theta_1)  dy_1 d \theta_1 \nonumber \\
&  = &  m \cdot  \int_{\theta_1 \notin B}  \int_{y_1 \in S_{marg}} \tilde{ \pi} (\theta_1) \cdot f( y_1 | \theta_1)  dy_1 d \theta_1
  =   E  \sum_{i=1}^m  \tilde{W}_i \nonumber
\end{eqnarray}
\end{pf}

\noindent
Notice that for $B = \emptyset$, $\sum_{i=1}^m W_i$ is the number of discoveries $R$.
While for $B = A_{marg}$, $\sum_{i=1}^m W_i$ is the number of false discoveries.
Therefore Lemma \ref{lemm0} implies that $E V$, $E R$, thus also  $pFDR = E V / E R$,
for $(\theta, Y)$ and for $(\tilde{\theta}, \tilde{Y})$ are the same.
According to Corollary \ref{cor0-a} the $pFDR$ for $(\tilde{\theta}, \tilde{Y})$ is the
corresponding ``random'' $\theta$ average risk, we denote $\tilde{r}_S$.
Thus since $FDR \le pFDR$, and $pFDR$ is the same
for $(\theta, Y)$ and for $(\tilde{\theta}, \tilde{Y})$, we get the following result.
\begin{cor} \label{cor00}
In the non-exchangeable random parameter model selecting $\theta_i$ if $y_i \in S_{marg}$
yields level $\tilde{r}_S$ $FDR$ control.
\end{cor}

To define a general method for specifying FDR controlling selection rules for $(\theta, Y)$
generated by the non-exchangeable random effect model with unknown marginal priors,
notice that applying empirical Bayes methods to $y_1 \cdots y_m$ actually estimates $\tilde{\pi} (\theta_i)$,
the mixture of the (unknown) marginal densities of $\theta_1 \cdots \theta_m$.
Combining this with Corollary \ref{cor00} implies that the FDR of any selection rule can be
approximated by $\tilde{r}_S$ computed by treating $(\theta, Y)$
as if it was generated by the random effect model and using the eBayes estimate of $\tilde{\pi} (\theta_i)$.
Furthermore, as $E R = E \tilde R$ and $E \tilde{R}  = m \cdot \Pr ( \tilde{Y}_i \in S_{marg})$,
then also in the non-exchangeable random effect model
the selection rule $S_{marg} = \{ y_i : \tilde{\rho} ( y_i) \le s \}$,
where $\tilde{\rho} (y_i)$ is the posterior expected loss in (\ref{re2ad}) computed for $(\tilde{Y}, \tilde{\theta})$,
yields the maximal $E R$ among all $S_{marg}$ with the same $\tilde{r}_S$.

\begin{dfn} \label{algo-FDR}
An algorithm for specifying level $q$ FDR controlling selection rules in
the non-exchangeable random effect model:
\begin{itemize}
\item [1.] Apply eBayes to $y_1 \cdots y_m$ to produce $\tilde{\pi} (\theta_i)$.
\item [2.] Use $\tilde{\pi} (\theta_i)$ to compute $\tilde{r}_S$ for any given selection rule.
\item [3a.] To specify a level $q$ FDR controlling selection rule of the form $S_{marg} = \{ y : T(y_i) \le s \}$,
for a given statistic $T(y_i)$, find $s$ for which $\tilde{r}_S = q$.
\item [3b.] The level $q$ FDR controlling selection rule yielding the maximal expected number of discoveries
is $S_{marg} = \{ y : \tilde{\rho} ( y_i) \le s \}$ with $s$, for which $\tilde{r}_S = q $.
\end{itemize}
\end{dfn}

\begin{exa} \label{exa444}
In Example \ref{exa1} selection is associated with $D = 2$ directional discoveries.
According to Corollary \ref{cor0-a} the pFDR
for the selection rule $|y_i| \ge s$ is equal to the ``random'' $\theta$
average risk  for the loss function $I( sign(\theta_i) \ne sign(y_i))$
\begin{equation} \label{eq1230}
E_{m_S (y)} \{  \ I ( y < -a) \cdot \Pr_{  \pi_S ( \theta | \; y )} ( \theta > 0) \;  + \;
 I ( y > a) \cdot \Pr_{ \pi_S ( \theta | \; y )} ( \theta < 0) \  \}.
\end{equation}
Recall that $|y_i| > 3.111$ was used to ensure that the directional-FDR is less than $0.1$.
For $s = 3.111$ the average risk (\ref{eq1230}) is $0.070$,
whereas setting $s = 2.915$ yields the selection criterion for which the average risk is $0.10$.
The posterior expected loss corresponding to the directional-FDR is
\[
\rho (y_i) = \Pr_{ \pi ( \theta | \; y )} ( sign(\theta_i) \ne sign(y_i)).
\]
Notice that in this example $\rho (y_i)$ increases in $|y_i|$,
thus $| y_i| \ge 2.915$ is the $r_S = 0.10$ selection rule yielding the maximal expected number of discoveries.
For $y_i \ge 0$, $\rho (y_i)$ is the conditional probability given $y_i$ that $\theta_i < 0$.
$\rho (0) = 0.5$, $\rho (3.111) =  0.176$, and $\rho (3.472) = 0.10$.
Thus $|y_i| \ge 3.472$ is the selection criterion suggested in (\ref{loc-FDR}) for $q = 0.10$.

The random effect model generated in Example \ref{exa1} is the $(\tilde{\theta}, \tilde{Y})$ that corresponds
to the non-exchangeable random effect model $(\theta, Y)$ in Example \ref{exa3333}.
To illustrate our results on the non-exchangeable random effect model,
we evaluated $E V$, $E R$ and the directional-FDR for $n = 10^5$ samples of
 $(\tilde{\theta}, \tilde{Y})$ and of $(\theta, Y)$.
In both cases the mean number of discoveries was $919.9$ (s.e. $<0.07$), the mean number of
false discoveries was $64.4$ (s.e. $<0.03$), and the mean directional-FDP was $0.070$ (s.e. $< 0.00003$).
\end{exa}

\section{The relation between saBayes inference and Bayesian FDR methods}
The term Bayesian FDR methods refers to the multiple testing procedures
presented in Efron et al. (2001) and Storey (2002, 2003)
for the following two group mixture model.
$H_i$, $i = 1 \cdots m$, are iid $Bernoulli(1-\pi_0)$ random variables.
$H_i = 0$ corresponds to a true null hypothesis, while $H_i = 1$ corresponds to a false null hypothesis.
Given $H_i = j$, $Y_i$ is independently drawn from $f_j$, for $j = 0, 1$.

The positive FDR (pFDR) corresponds to a rejection region $\Gamma$.
It is defined $E ( V / R | \; R >0)$ where $R$ is the number of $y_i \in \Gamma$,
and $V$ is the number of $y_i \in \Gamma$ with $H_i = 0$.
Storey proves that
\begin{eqnarray}
\lefteqn{ pFDR ( \Gamma )
 =    Pr( H_i = 0 | Y_i \in \Gamma )} \label{pFDR-1} \\
& = &  \frac{ \pi_0 \cdot Pr( Y_i \in \Gamma  | H_i = 0 )}
{ \pi_0 \cdot Pr( Y_i \in \Gamma  | Y_i = 0 ) + (1-\pi_0) \cdot Pr( Y_i \in \Gamma | H_i = 1  )}, \label{pFDR-2}
\end{eqnarray}
with $Pr( Y_i \in \Gamma  | H_i = j ) = \int_{y_i \in \Gamma} f_j(y_i) dy_i$.
For the multiple testing procedure each null hypothesis is associated with a rejection region $\Gamma_i$,
determined by $y_i$;
the pFDR corresponding to $\Gamma_i$, called the q-value, is computed;
and the null hypothesis $H_i = 0$ is rejected if $\hbox{q-value} \le q$.
The local FDR is defined in Efron et al. (2001)
as the conditional probability given $Y_i = y_i$ that $H_i = 0$
\[
fdr ( y_i )  =   \frac{ \pi_0 \cdot f_0 (y_i) }
{ \pi_0 \cdot f_0 (y_i)  + (1-\pi_0) \cdot f_1 (y_i)}.
\]
The multiple testing procedure based on the local FDR is to reject $H_i = 0$ if $fdr(y_i) \le q$.

Notice that Bayesian FDR methods can be expressed as a special case of the FDR controlling selection
rules presented in the previous section, in which the components of the parameter vector are dichotomous.
The parameter is $H = (H_1 \cdots H_m)$, and $(H, Y)$ are generated by a random effect model:
the marginal distribution of $H_i$ is $\pi (H_i = j) = \; (1 - \pi_0)^j \cdot \pi_0^{(1-j)}$,
$f_j$ is the likelihood, $H_i$ is selected if $y_i \in \Gamma$ and selection is associated
with the discovery that $H_i = 1$.
Notice also that Expression (\ref{pFDR-2}) is a special case of Expression (\ref{pFDR=saprior}):
it is the ``random'' parameter average risk for the loss function $I( H_i = 0)$,
expressed as the selection-adjusted prior distribution of making a a false discovery
\[
\pi_{\Gamma} ( H_i = 0 )  \; \propto  \; \pi(H_i = 0) \cdot \Pr ( Y_i \in \Gamma | \; H_i = 0 ).
\]
Thus the equality in (\ref{pFDR-1}) proven by Storey is a special case of Corollary \ref{cor0}.
The local FDR is the ``random'' $\theta$ selection-adjusted posterior expected loss,
thus the multiple testing procedure based on the local FDR is a special case of the
selection rule in (\ref{loc-FDR}).
Lastly, the relation between the local FDR and the pFDR, $pFDR  = E_{y \in \Gamma}  fdr(y)$,
follows from the definition of the average risk in (\ref{saBayes-risk}).

Bayesian FDR methods are valid regardless of whether $H$ is a ``random'' or ``fixed'' parameter.
However in selective inference for $h_i(H) = H_i$,
the selection-adjusted posterior probability that $H_i = 0$ for a ``random'' $H$ is equal to the local fdr.
Whereas if $H$ is a ``fixed'' parameter, or if $\pi_{0}$ is the non-informative prior probability that $H_i = 0$,
then the selection-adjusted posterior distribution that $H_i = 0$ is
\[
 \frac{  \pi_{0}\cdot f_{\Gamma} ( y_i  | H_i = 0 )  }
{ \pi_{0} \cdot f_{\Gamma} ( y_i | H_i = 0) + (1 - \pi_{0} ) \cdot f_{\Gamma} ( y_i | H_i = 1 )},
\]
for $f_\Gamma (y_i | \; H_i = j) = f_j (y_i) / \Pr ( y_i \in \Gamma | \; H_i = j )$
the selection-adjusted likelihood.

\section{Analysis of microarray data}
We analyze the Dudoit and Yang (2003) swirl data set.
The data includes $4$ arrays with $8448$ genes, comparing RNA from Zebrafish with the swirl mutation to RNA from wild-type fish.
For Gene $g$, $g = 1 \cdots 8448$, the parameters are $\mu_g$ the expected log2-fold change in expression
due to the swirl mutation, and $\sigma^2_g$ the variance of the log2-fold change in expression.

In our analysis we assume that $(\theta, Y)$ are generated by a non-exchangeable random effect model.
Since the measurement error variances are expected to vary from experiment to experiment,
$\sigma^2_g$ are iid ``random'' parameters with scaled inverse chi-square marginal prior density
$\pi( \sigma^2_g)$, whose hyper-parameters, $s^2_0 = 0.052$ and $\nu_0 = 4.02$,
were derived by applying the R LIMMA package (Smyth, 2005) {\em eBayes}
function to the sample variances.
While $\mu_g$ are distinct independent ``fixed'' parameters
that are elicited flat non-informative priors, $\pi_{ni} (\mu_g) \propto 1$.
However for assessing the FDR of the
BH procedure and for specifying the Bayesian selection rules we use the eBayes prior
\[
\tilde{\pi} (\mu_g) \; = \; 8.5  \cdot \exp ( - 8.5 \cdot | \mu_g|) / 2,
\]
that provided a good fit to the empirical distribution of $\bar{y}_1 \cdots \bar{y}_{8448}$.
Given $\mu_g$ and $\sigma_g$,
$s^2_g$ the sample variances  are independent $\sigma^2_g \chi^2_{3} / 3$,
and $\bar{y}_g$ the observed mean log2 expression ratios are independent $N( \mu_g , \; \sigma^2_g / 4)$.
Thus the marginal likelihood is given by
\begin{equation} \label{eqG}
f ( \bar{y}_{g}, s^2_{g} | \; \mu_{g}, \sigma^2_{g} ) \; \propto \;
\sigma_{g}^{-4} \exp \{ - \frac{1}{2 \sigma_{g}^2} [
3 s_{g}^2 + 4 ( \mu_{g} - \bar{y}_{g})^2 ] \}.
\end{equation}

Our goal in the analysis is to specify a selection rule for which the mean directional error
in declaring selected genes with $\bar{y}_g > 0$ over-expressed
and declaring selected genes with $\bar{y}_g < 0$ under-expressed is less than $0.05$,
and to provide inference for the change in expression of selected genes.

\subsection{Specifying the selection rules}
In the first part of our analysis we apply the level $q = 0.10$ BH procedure to moderated
t-statistic p-values to
discover differentially expressed genes;
assess the directional-FDR of the selection rule specified by the BH procedure;
and compare its performance to
the performance of the level $q = 0.05$ directional FDR controlling selection rules based on
moderated $t$ statistics and on the posterior expected loss.

LIMMA implements a hybrid classical/Bayes approach in which $\mu_g$ are assumed to be unknown constants
while $\sigma^2_g$ are iid $\pi(\sigma^2_g)$.
The moderated $t$ statistics are defined $\tilde{t}_g = \bar{y}_g  / (\tilde{s}_g  / 2)$,
for $\tilde{s}^2_g =  (\nu_0 s^2_0 + 3 s^2_g)/(\nu_0 + 3)$
the posterior mean of $\sigma^2_g | s^2_g$.
As $\tilde{s}^2_g / \sigma^2_g \sim \chi^2_{\nu_0 + 3} / (\nu_0 + 3)$,
$(\bar{y}_g - \mu_g)  / (\tilde{s}_g  / 2)$ are $(\nu_0 + 3)$ degrees of freedom $t$ random variables.
Thus the p-values LIMMA provides to test a null hypothesis of non-differential expression is
$\tilde{p}_g = 2 \cdot (1- F_{\nu_0 + 3} (|\tilde{t}_g|))$, where $F_{\nu}$ is the $\nu$ degrees of freedom $t$ cdf.
Applied at level $q = 0.10$ to the $8448$ p-values the BH procedure yielded $245$ discoveries,
corresponding to the rejection region $|\tilde{t}_g| >  4.479$.
The observed mean log2 expression ratios and sample standard deviations of the $8448$ genes are drawn in Figure 5.
The BH discoveries are the $245$ observations beneath the solid blue curve $|\tilde{t}_g| =  4.479$.
To see why this rejection region corresponds to $0.05$ directional FDR control
notice that for all $\mu_g$, the probability of a directional error is less than $1- F_{\nu_0 + 3} (4.479)$ ;
thus $12.08 = 8448 \cdot (1- F_{\nu_0 + 3} (4.479))$ is a conservative estimate for the number of false directional discoveries,
and $0.049 = 12.08 / 245$ is a conservative estimate for the directional FDR.

For comparison, the frequentist treatment of this problem would be to test
the null hypotheses of non-differential expression by 3 degrees of freedom test statistics $t_g = \bar{y}_g  / (s_g  / 2)$.
Since the 3 degrees of freedom t-distribution has heavier tails,
$F^{-1}_3 ( 1 - 0.1 / (2 \cdot 8448)) = 57.10$ while $max (| t_g|)$ is only $27.90$.
Thus applying the level $q=0.1$ BH to $p_1 \cdots p_{8448}$, with $p_g = 2 \cdot (1- F_{3} (|t_g|))$,
yields no discoveries.

In order to assess the directional FDR we derive the ``random'' $\theta$ saBayes posterior distribution
\begin{equation} \label{saBayes-eBayes}
\tilde{\pi}_S (  \mu_{g}, \sigma^2_g | \bar{y}_{g}, s_{g})  \; = \; \frac{
I ( (\bar{y}_g, s^2_g) \in S_{marg} ) \cdot
\tilde{\pi} (\mu_g, \sigma^2_g) \cdot f ( \bar{y}_{g}, s_{g} | \; \mu_{g}, \sigma^2_{g} )}
{\Pr(  (\bar{y}_g, s^2_g) \in S_{marg})},
\end{equation}
for the eBayes prior distribution $\tilde{\pi}( \mu_g, \sigma^2_g) = \tilde{\pi}(\mu_g)  \cdot \pi (\sigma^2_g)$.
We then integrate out $\sigma^2_g$ in (\ref{saBayes-eBayes}) to derive
$\tilde{\pi}_S (  \mu_{g} | \bar{y}_{g}, s_{g})$ the marginal ``random'' $\theta$ saBayes posterior distribution of $\mu_g$,
and the ``random'' $\theta$ posterior expected loss corresponding to directional errors
\[
\tilde{\rho} (\bar{y}_{g}, s^2_{g}) \; = \; \int I \{  \mu_g \ne sign(\bar{y}_{g}) \}   \cdot \tilde{\pi}_S (  \mu_{g} | \bar{y}_{g}, s^2_{g}  \;)
d \mu_g,
\]
and use it to numerically compute the ``random'' $\theta$ average risk corresponding to the directional FDR
\[
\tilde{r}_S ( S_{marg}) \; = \; E_{m_S (\bar{y}_{g}, s^2_{g})} ( \tilde{\rho} (\bar{y}_{g}, s^2_{g}) ),
\]
for
\[
m_S (\bar{y}_{g}, s_{g}) \;  = \; \frac{I ( (\bar{y}_g, s^2_g) \in S_{marg} ) \cdot
\tilde{\pi} (\mu_g, \sigma^2_g) \cdot f ( \bar{y}_{g}, s_{g} | \; \mu_{g}, \sigma_{g} )  }
{ \int I  (   (\bar{y}_g, s^2_g) \in S_{marg}) \cdot  \tilde{\pi} (\mu_g, \sigma^2_g)
 \cdot
f ( \bar{y}_{g}, s_{g} | \; \mu_{g}, \sigma_{g} )d \mu_{g} d\sigma_{g}}.
\]
$\tilde{r}_S$ for $| \tilde{t}_g| > 4.479$ the $q = 0.10$ BH procedure (solid blue curve in Figure 5)
is $0.024$.
While $|\tilde{t}_g| >  2.64$ (dashed blue curve in Figure 5)
is the moderated $t$ selection rule with $\tilde{r}_S = 0.05$. It yields $1124$ discoveries.
The green curves in Figure 5 correspond to the selection rules of the form $\tilde{\rho} (\bar{y}_{g}, s^2_{g}) < s$.
The solid curve corresponds to the selection rule with $s = 0.05$, that yields $559$ discoveries.
The dashed curve corresponds to the selection rule with $s = 0.088$, for which $\tilde{r}_S = 0.05$.
This is the selection rule that yields the maximal expected number of discoveries among all selection rules with $\tilde{r}_S = 0.05$.
In this case it yields $1271$ discoveries.

\subsection{Providing saBayes inference}
In the second part of our analysis we provide saBayes inference for $\mu_{6239}$,
the expected log2-fold change in expression due to the swirl mutation for Gene number $6239$.
The statistics for this gene (marked by the red plus sign in Figure 5)
are $\bar{y}_{6239} = -0.435$ and $s^2_{6239} = 0.0173$, thus $\tilde{t}_{6239} = -4.51$.
Note that a frequentist solution to this problem would be to construct a
FCR adjusted, 3 degrees of freedom $t$ distribution, marginal confidence interval for $\mu_{6239}$.

The marginal posterior distributions of $\mu_{6239}$ are drawn in Figure 6.
The black curve corresponds to the non-informative prior unadjusted posterior
\[
\pi (  \mu_{g}, \sigma^2_{g} | \bar{y}_{g}, s^2_{g}  \;) \; \propto \;
 \pi_{ni} (\mu_g) \cdot \pi (\sigma^2_g)\cdot f ( \bar{y}_{g}, s_{g} | \; \mu_{g}, \sigma^2_{g} ),
\]
for which $( \mu_{6239} - \bar{y}_{6239}) / (\tilde{s}_{6239} / 2) \sim t_{7.02}$.
In this case, the posterior mean and mode equal $\bar{y}_{6239} = -0.435$,
the $0.95$ credible interval for $\mu_{6239}$ is $[ -0.61, -0.21 ]$,
the posterior probability that $\mu_{6239} > 0$ and a directional error is committed is $0.0014$.
The green curve corresponds to $\tilde{\pi}_S (  \tilde{\mu}_{6239} | \bar{y}_{6239}, s_{6239}  \;)$.
Its posterior mode is $-0.36$, the posterior mean is $-0.31$, the $0.95$ credible interval is $[ -0.54, -0.01]$,
and the posterior probability that  $\mu_{6239} > 0$ is $0.020$.

As $\mu_g$ is elicited a non-informative prior and $\sigma^2_g$ is a ``random'' parameter,
then $(\mu_g, \sigma^2_g)$  is a ``mixed'' parameter,
and its selection-adjusted posterior distribution is proportional to the
joint truncated distribution in (\ref{dens-lambda}), with $\mu_g$ substituting the ``fixed'' $\lambda$
and $\sigma^2_g$ substituting the ``random'' $\theta$,
\begin{eqnarray} \label{eq44}
\lefteqn{\pi_S (  \mu_{g}, \sigma^2_{g} | \bar{y}_{g}, s^2_{g}  \;) \propto
f_S (  \mu_{g}, \sigma^2_{g},  \bar{y}_{g}, s^2_{g}  \;)} \\
& = &  \pi (\sigma^2_g) \cdot \pi_{ni} (\mu_g) \cdot f ( \bar{y}_{g}, s^2_{g} | \; \mu_{g}, \sigma^2_{g} )
 / \Pr( |\tilde{t}_g| > a \ | \; \mu_g ). \nonumber
\end{eqnarray}
SaBayes inference for $\mu_{6239}$ is based on $\pi_S (  \mu_{g} | \bar{y}_{g}, s_{g})$,
the marginal selection adjusted posterior of $\mu_{6239}$, derived by integrating out $\sigma^2_g$ from (\ref{eq44}).
The solid blue curve is $\pi_S (  \mu_{g} | \bar{y}_{g}, s^2_{g}  \;)$
for the selection rule $|\tilde{t}_g| > 4.479$.
Its posterior mode is $-0.278$, the posterior mean is $-0.257$, the $0.95$ credible interval is $[ -0.54, 0.02]$,
and the posterior probability that $\mu_{6239} > 0$, and thus the Gene was erroneously declared under-expressed, is $0.038$.
The dashed blue curve corresponds to $|\tilde{t}_g| > 2.64$.
In this case the shrinking towards $0$ is weaker:
the posterior mode is $-0.419$, the posterior mean is $-0.367$, the $0.95$ credible interval is $[ -0.63, -0.02]$,
and the posterior probability that $\mu_{6239} > 0$ is $0.017$.

\section{Discussion}

The observation that selection affects Bayesian inference carries the important implication that in
Bayesian analysis of large data sets, for each potential parameter, it is necessary to explicitly specify
a selection rule that determines when inference is provided for the parameter
and provide inference that is based on the selection-adjusted posterior distribution of the parameter.

Even though specifying a selection rule
introduces an arbitrary element to Bayesian analysis,
it is important to note that the selection rule is determined before the data is observed,
and once the selection rule is determined the entire process of providing saBayes inference
is fully specified and is carried out the same way as Bayesian inference.
The notable exception is eBayes methods that use the data twice in the analysis,
first to elicit the prior distribution and possibly to specify the selection rule,
and then to produce posterior distributions.

Our method of controlling the Bayesian FDR corresponds to the
fixed rejection region approach presented in Yekutieli and Benjamini (1999),
that consists of estimating the FDR in a series of nested fixed rejection regions and choosing
the largest rejection region with estimated FDR less than $q$.
However, as the pFDR of any selection rule can be expressed as a saBayes risk,
the problem of controlling the Bayesian FDR in the random effect and non-exchangeable random effect models
is reduced into a Bayesian decision problem of finding the ``optimal''
selection rule with saBayes risk $\le q$.
Our Bayesian FDR controlling methods can, in principle,  provide tight FDR control, based on the ``optimal'' statistic,
for any discovery event.
Whereas frequentist FDR controlling methods may provide tight FDR control when the discovery is rejecting a simple null hypothesis,
but as illustrated by the performance of the BH procedure in controlling the directional-FDR,
can only bound the FDR when the discoveries are rejecting composite null hypotheses.

In general, the price paid by using stricter selection rules is
reduction in the information the data provides for selective inference.
Example \ref{ex89} suggests that when specifying selection rules, in addition to
the tradeoff between allowing too many false (or wasteful) discoveries and failing to make enough discoveries,
it may also be advisable to take into account the quality of the inference provided for
selected parameters.


\begin{figure} \label{fig1}
\includegraphics[scale=.8,angle=0]{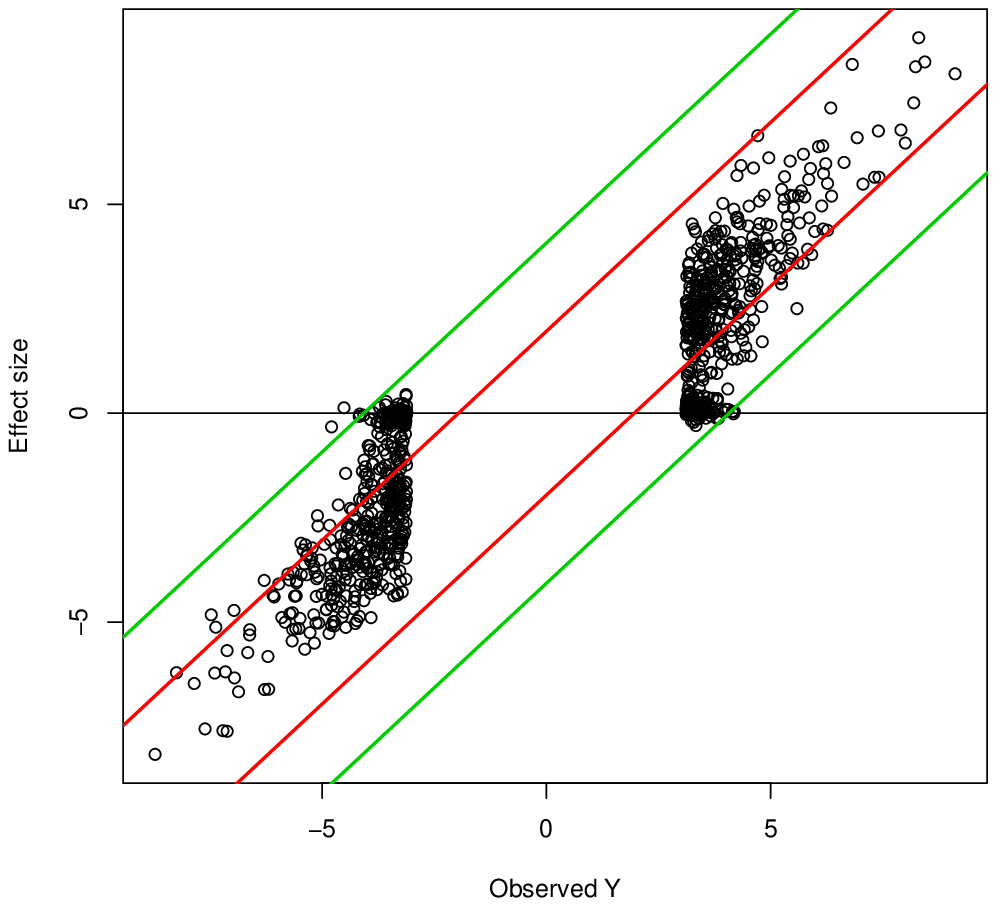}
\caption{Simulated example -- scatter plot of $|Y_i| > 3.111$ components.
$Y_i$ values are drawn on the abscissa of the plot, the ordinates are $\theta_i$ values.
The red lines are marginal $0.95$ CIs.
The green lines are $0.05$ FCR-adjusted CIs.}
\end{figure}

\begin{figure} \label{fig3a}
\includegraphics[scale=.8,angle=0]{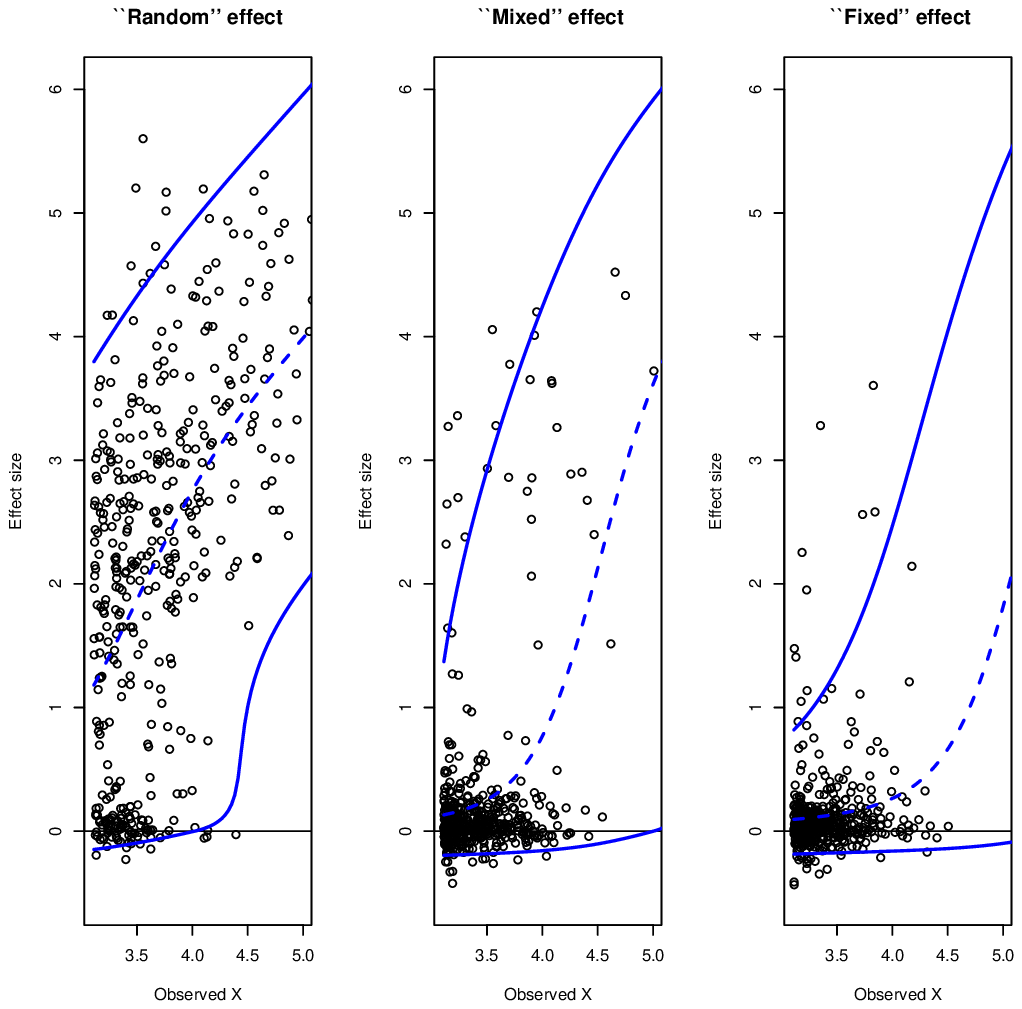}
\caption{Simulated example -- scatter plot of $Y_1 > 3.111$ realizations of $(\theta_1, Y_1)$ in
the ''random'' parameter truncated sampling model (left panel -- $466$ observations),
the ''mixed'' parameter truncated sampling model (middle panel -- $498$ observations),
and the ''fixed'' parameter truncated sampling model (right panel -- $501$ observations).
The solid blue curves are the selection-adjusted $0.95$ posterior credible intervals for $\theta_1$,
and the dashed blue curves are the selection-adjusted posterior means.}
\end{figure}

\begin{figure} \label{fig2}
\includegraphics[scale=.8,angle=0]{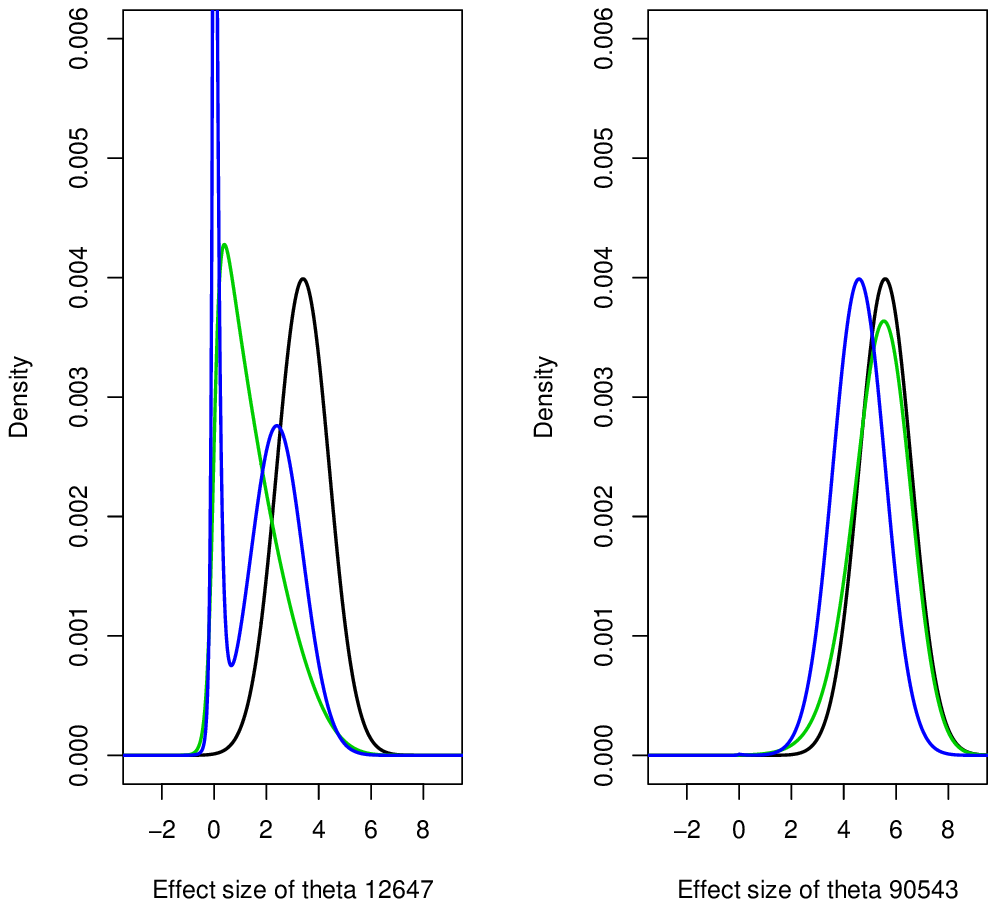}
\caption{Simulated example -- saBayes posterior distributions.
The Posterior distributions for $\theta_{12647}$ are drawn in the left panel,
the Posterior distributions for $\theta_{90543}$ are drawn in the right panel.
The black curves are unadjusted posteriors;
the blue curves are ``random'' parameter model saBayes posteriors;
the green curves are non-informative prior saBayes posteriors.}
\end{figure}

\begin{figure} \label{fig3}
\includegraphics[scale=.8,angle=0]{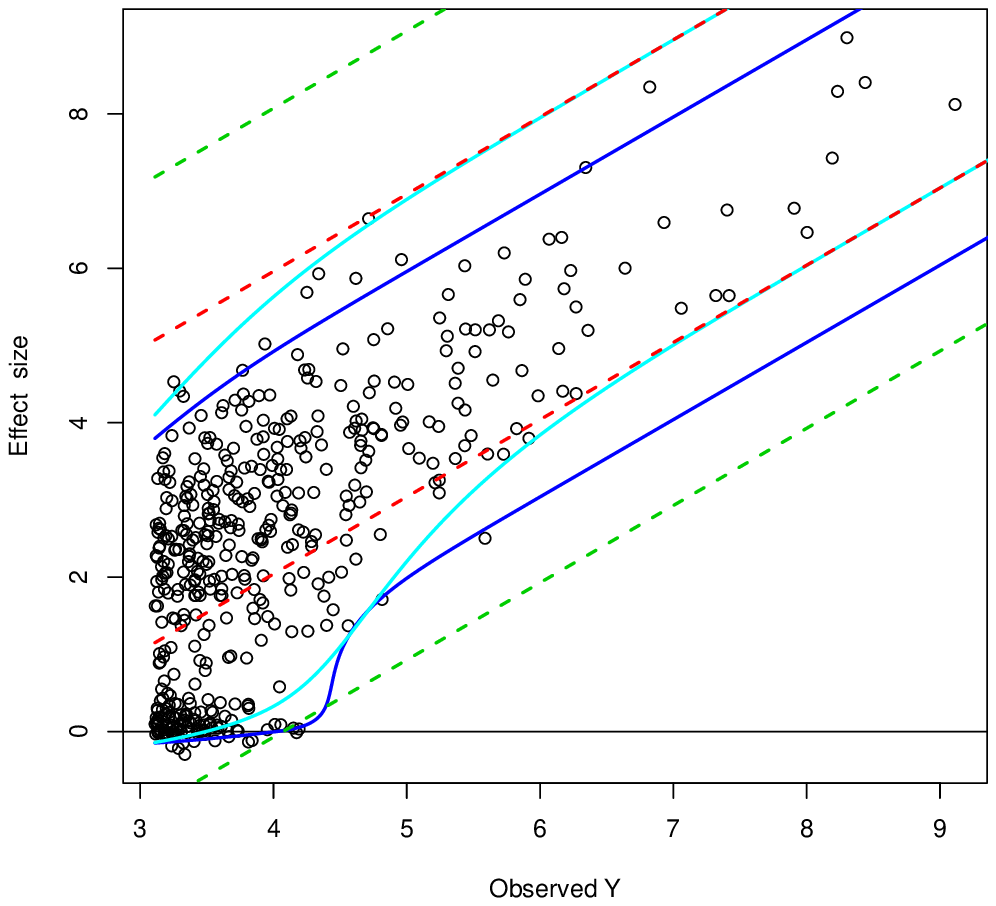}
\caption{Simulated example -- scatter plot of $Y_i > 3.111$ components.
The dashed green and red lines are the CIs from Figure $1$.
The blue curves are the ``random'' parameter model saBayes $0.95$ credible intervals.
The light-blue curves are the non-informative prior saBayes $0.95$ credible intervals.}
\end{figure}

\begin{figure} \label{fig7a}
\includegraphics[scale=.8,angle=0]{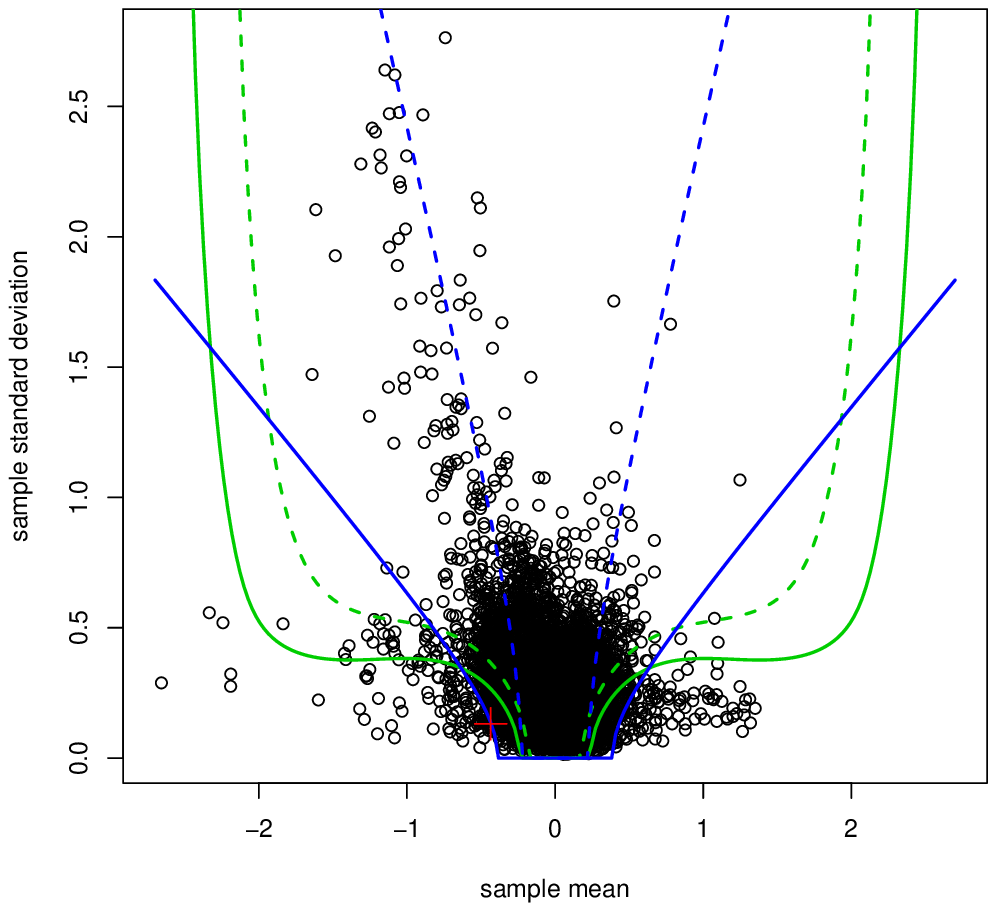}
\caption{Swirl data -- scatter plot of sample means and standard deviations.
The abscissa of the plot is $\bar{y}_g$, the ordinates are $s_g$.
The solid blue curve is $|\tilde{t}_g| = 4.479$.
The dashed blue curve is $|\tilde{t}_g| = 2.64$.
The solid green curve is $\tilde{\rho}(\bar{y}_{g}, s_{g}) = 0.05$.
The dashed green curve is $\tilde{\rho}(\bar{y}_{g}, s_{g}) = 0.088$.
The red plus sign is $(\bar{y}_{6239}, s_{6239})$.}
\end{figure}

\begin{figure} \label{fig7b}
\includegraphics[scale=.8,angle=0]{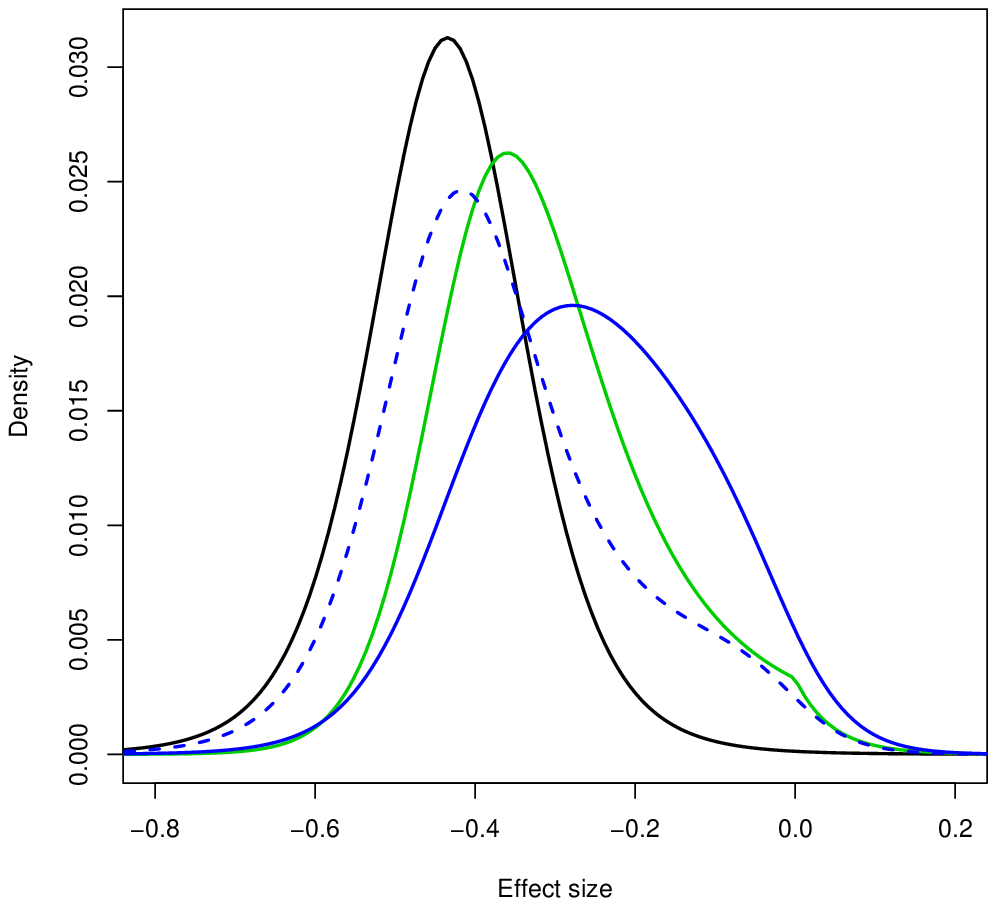}
\caption{Swirl data -- marginal posterior densities of $\mu_{6239}$.
The black curve is the non informative prior unadjusted posterior distribution.
The green curve is the eBayes prior posterior distribution.
The solid blue curve is the non-informative prior saBayes posterior distribution for the selection rule $|\tilde{t}_g| > 4.479$.
The dashed blue is the non-informative prior saBayes posterior distribution for the selection rule$|\tilde{t}_g| > 2.64$.}
\end{figure}

\end{document}